\documentclass[11pt,a4paper]{article}
\usepackage[utf8]{inputenc}
\usepackage{amsfonts}
\usepackage{amssymb}
\usepackage{dsfont}
\usepackage{amsmath} 
\usepackage{amsthm} 
\usepackage{enumerate}
\usepackage{pdfpages}
\usepackage{cite} 
\usepackage{tikz} 
\usetikzlibrary{shapes,arrows}
\usepackage{float}\usepackage{wrapfig}
\usepackage{hyperref} 
\hypersetup{
    colorlinks=true,
    linkbordercolor = {white},
    citecolor= {blue},
    linkcolor = .,
}
\usepackage{subcaption} 
\usepackage{color}
\usepackage{amsmath}
\usepackage{amsfonts}
\usepackage{amssymb}
\usepackage{graphicx}
\usepackage[left=2cm,right=2cm,top=2cm,bottom=2cm]{geometry}
\begin{document}
\title{\textbf{A complex network framework to model cognition: unveiling correlation structures from connectivity}}

\author{\sc Gemma Rosell-Tarrag\'{o}$^*$, Emanuele Cozzo and Albert D\'{i}az-Guilera\\
\small{Departament de F\'{i}sica de la Mat\`{e}ria Condensada, Universitat de Barcelona, Barcelona, Spain} \\
\small{Universitat de Barcelona Institute of Complex Systems (UBICS), Universitat de Barcelona, Barcelona, Spain}
\\\small{$^*$gemma.rosell.tarrago@ub.edu}}

\date{}
\maketitle
\begin{abstract}
Several approaches to cognition and intelligence research rely on statistics-based models testing, namely factor analysis. In the present work we exploit the emerging dynamical systems perspective putting the focus on the role of the network topology underlying the relationships between cognitive processes. We go through a couple of models of distinct cognitive phenomena and yet find the conditions for them to be mathematically equivalent. We find a non-trivial attractor of the system that corresponds to the exact definition of a well-known network centrality and hence stress the interplay between the dynamics and the underlying network connectivity, showing that both of the two are relevant. The connectivity structure between cognitive processes is not known but yet it is not any. Regardless of the network considered, it is always possible to recover a positive manifold of correlations. However, we show that different network topologies lead to different plausible statistical models concerning correlations structure, ranging from one to multiple factors models and richer correlation structures. 
\end{abstract}

\section{Introduction}
\label{section:introduction}

Individuals differ from one another in their ability to learn from experience, to adapt to new situations and overcome challenges, to understand simple to complex ideas, to solve real-world and abstract problems and to engage in different forms of reasoning and thinking. Such differences in performance occur even in the same person, in different domains, across time and distinct yardsticks \cite{Tucker-Drob2010,Kanfer1990,Jonassen1993,Sternberg}. 

These complex cognitive processes are intended to be clarified and put together by the concept of intelligence. Despite many advances have been made, there are still open questions regarding its building blocks and nature yet to be solved \cite{Neisser1996,Smith2009,Eysenck1988}.

There is a fair amount of research carried out and still going on about the theory of intelligence and few statements have been unequivocally been established. Furthermore, it should be noted the easiness that ongoing research gives rise to society policies, despite prevailing disputes and great unknowns are present.\cite{Herrnstein1996,Smith2009,Rindermann2007}. For this reason, there is an urgent need to understand the most important root causes, validate existing theories and shed light to people who are responsible of educational and even social and health decision-making.

Nowadays, there is a significant number of approaches to intelligence. Developmental psychologists are often more concerned about intelligence as a subset of evolving processes throughout life, rather than about individual differences \cite{Zachary2009}. Several theorists stress the role of culture in the very conceptualization of intelligence and its influence in individuals \cite{Lesser1965}, while others point to the existence of different intelligences, either measurable or not \cite{Gardner1983}. There is also an increasing interest in contributions coming from biology and neuroscience \cite{Erlenmeyer-Kimling1963,Deary2010,Gray2004,Mattar2016,Donovan}. Yet, the most influential approach so far is based on psychometric testing \cite{Carroll1993,Ackerman1987,Kaufman2012,Castejon2010,Molenaar2010,Johnson2008,Ackerman2005}

Psychometrics has enabled successful and systematic measures of a wide range of cognitive abilities like verbal, visual-spatial, fluid reasoning, working memory, processing speed among others through standardized tests \cite{Wechsler1939,Terman1960}. Even if distinct, these assessed abilities turn out to be intercorrelated rather than autonomous prowesses. That is, people who perform well in a given test tend to obtain higher scores on the others as well. This well-documented evidence concerning positive correlations between tests, regardless of its nature, is called the positive manifold. And precisely because of the existence of such complex relations, one of the main aim of this approach is to unveil the structure which best describes the relationships between a number of distinguishable factors or aptitudes that may exist. On this basis, many studies use exploratory and confirmatory factor analysis techniques, starting off from between-tests correlation matrices.

Furthermore, there exists a complex correlation structure between abilities which may unveil the underlying connection between cognitive processes. Factor analysis might help clearing up such patterns and yet bring about discussion on the meaning of the outcome.

A brief historical overview since the early days of intelligence research and its development may help us understand the spectrum of existing models. Some theorists relied on the shared variance among abilities, which Charles Spearman, pioneer of factor analysis, called the \textit{g} factor or general intelligence \cite{Spearman1904}, i.e one common factor which explains most of the variance within a population and source of improvement or decline of all other abilities, and it is still cause for controversial. 

Alternatively, hierarchical models of intelligence where each layer accounts for the variations in the correlations within the previous one were also well accepted. \cite{Cattell1963,Horn1966,Carroll1993}.

Nevertheless, a fairly number of scholars argued against theories of cognitive abilities or intelligence drawn upon the concept, measure and meaning of general intelligence. Namely, Howard Gardner, stated that an individual has a number of relatively autonomous intellectual capacities, with a degree of correlation empirically yet to be determined, called multiple intelligences, among which non-cognitive abilities are included  \cite{Gardner1983}. 

Two different approaches with reference to the relationship between observable variables and attributes or constructs prevail in present research and theorizing in psychology, but also clinical psychology, sociology and business research amongst others: formative and reflective models \cite{Schmittmann2013}. In the first of this conceptualization, observed scores define the attribute, whereas in the latter, the attribute is considered as the common cause of all observables. As an example, the classic definition of general intelligence could fall into a reflective model. But also, in clinical psychology, a mental disorder may be thought to be a reflective construct that brings about its observable symptoms \cite{Dolan2009}. Possible correlation between observables might be therefore due to its underlying common cause. Conversely, the aggregate outcome of education, job, neighbourhood and salary leads to socio-economic status (SES), a standard example of formative model.

A more recent approach aims to combine distinct possible factor models by only using the information about the factorial structure found by each study.\cite{Massidda2016}. 
 
Both formative and reflective models, along with similar alternatives, may elicit discussion regarding two different issues: one first source of debate is rooted in the meaning and interpretation of such models, while a second cause stems from disregarding the role of time, that is, the dynamics of the system is not explicitly considered.

The above-mentioned problems can potentially be overcome if we consider that variables, i.e, observables, scores or indicators, are the characteristics of nodes in a network. These latter are directly connected through edges, which reflect the coupling between variables. Dynamical systems theory is therefore the proper framework to formalize and study the behaviour of such systems \cite{Juarrero2010}. Starting from an initial state, the system evolves in time according to a system of coupled differential equations and eventually reaches an attractor state of the system.

Noteworthy, a substantive piece is prevalently missing: the topology of the network on where the process is taking place may be a determinant fact which enables nodes to communicate between each other and brings about correlations not explicitly enforced in the model. Therefore, the objective and contribution of the present work is exploring the significant role of network topology or connectivity structure between the variables deemed meaningful to the case of cognitive abilities or intelligence models.

In this work we evince the tight connection between a centrality measure of the network and the stable solution of the studied models. Moreover, we show that distinct network topologies may explain different correlation structures.

The paper is organized as follows: Section \ref{section:network} introduces basic notions of networks and explored topologies. Section \ref{section:cognition} describes and formalizes the two studied models of cognition. Section \ref{section:interplay} and Section \ref{section:correlations} go through the main results, concerning dynamics and correlations, respectively. Final discussion and conclusions are presented in last section. Further mathematical methods can be found in the Appendix. 

\section{Network topology}
\label{section:network}
A network, $G(V,E)$, is a collection of vertices or nodes, $V(G)$, linked by edges, $E(G)$, which are given meaning and attributes. Networks can describe complex interconnected systems such as social relationships, transportation maps, economic, biological and ecological systems among others. We consider networks that have neither self-edges nor multiedges, called \textit{simple networks}\cite{Newman2010}.

The adjacency matrix of $G$, written $A(G)$, is the $N$-by-$N$ matrix which entry $A_{ij}$ equals $1$ if node $i$ is linked to node $j$, and $0$ otherwise. Networks can be directed or undirected, although we stay on the latter case. 

The topology of a network characterizes its shape or structure and the distribution of connections between nodes. Besides the attributes of nodes and edges, the topology of a network determines its main properties and makes it distinguishable from others. One main property is the degree of a node $i$, $k_i$, which is the number of edges connected to it. Although networks may describe particular real systems, regardless of its nature, they can be classified to one of the most well-known families of networks. Right after we briefly describe the four network models explored in the present work.

\begin{enumerate}[(a)]
\item Complete network
\label{item:complete}

Within the family of deterministic networks, a complete network is characterized by its nodes being fully connected, that is, each node is connected to the others, such that all off-diagonal elements of the adjacency matrix are equal to $1$, $A_{ij}=1 \ \forall i \neq j$. 

\item Erd\"{o}s-R\'{e}nyi network
\label{item:erdos}

One of the most renowned random network is generated by the Erd\"{o}s-R\'{e}nyi (ER) model \cite{Erdos1959}. Given the number of nodes, $N$, and the probability of an edge, $p$, this model, $G(N,p)$, chooses each of the possible edges with probability $p$.
However, generally, real networks are better described by heterogeneous rather than ER networks. Therefore, ER networks are often used as null hypothesis to reject or accept models concerning more complex situations. 
\item Heterogeneous network
\label{item:heterogeneous}

There is a wide range of networks coming from real systems (either found in nature or human driven) which topology is far from being homogeneous, but it rather entails degree distributions which are characterized by a power law, also called scale-free when the networks are large enough.\cite{Barabasi2009}. The Internet network, protein regulatory networks, research collaborations, on-line social networks, airline systems, cellular metabolism, companies and industries interlinks are few examples of them \cite{Caldarelli2013,Albert2002} .

\item Newman modular network
\label{item:newman}

In addition to the degree distribution, another important feature is the presence of communities or modules within a network, mainly in social, but also in metabolic or economic networks \cite{Ravasz2002,Girvan}. A module or community can be defined as a subset of nodes which is more densely linked within it than with other subsets of nodes. 

One particular method to generate such modules within a network is the Newman model, which distributes the nodes in a number, $N_{modules}$, of modules not necessarily isolated from the others. \cite{Danon2005a,Newman,Newmana}. Similarly as ER networks, with a probability $p_{in}$, an edge between pairs of nodes belonging to the same community is created, whereas pairs belonging to different communities are linked with probability $p_{out}$. In the model, the number of nodes, $N$, the total average degree, $\left\langle k \right\rangle$ and $ \left\langle k_{in} \right\rangle$, which stands for the average degree within a community, are fixed. Hence, $p_{in}$ and $p_{out}$ are given by:
\begin{eqnarray}
p_{in}=\frac{\left\langle k_{in} \right\rangle}{n_{in}-1} & & p_{out}=\frac{\left\langle k \right\rangle - \left\langle k_{in} \right\rangle}{n_{out}}
\end{eqnarray}
where $n_{in} \equiv N/N_{modules}$ and $n_{out} \equiv N - n_{in}$.

As $\left\langle k_{in} \right\rangle$ grows the network modularity increases \cite{Newman2006}, that is, the communities become easier to identify. 
\end{enumerate}

\section{Models of cognition}
\label{section:cognition}
Within the framework of dynamical systems and network theory, there exists a one-to-one map between variables and nodes, such that variable $i$ is represented by node $i$. The value of variable $i$, $x_i$ is set as an attribute of its corresponding node. In this new space, the adjacency matrix, which maps the interactions between variables on a network, can lump exogenous effects together in a very compact way: 
\begin{equation}
\label{general}
\dot{x_i}(t) = F_i(x_i,t)+\sum_{j}A_{ij}(t)G_{ij}(x_i, x_j,t)
\end{equation}
Therefore, (\ref{general}) is the most general expression which integration determines the temporal evolution of each variable, $x_i(t)$.
$F_i(x_i,t)$ accounts for endogenous effects, i.e, a function that depends only on variable $x_i$. $G_{ij}(x_i, x_j, t)$ takes into account exogenous effects on $i$, i.e, a function that describes the influence of its neighbouring variables, $x_j$. The intensity of such individual interactions are included in $G_{ij}$ in the form of weights. The adjacency matrix, $A$, determines whether variables are coupled between them: if variables $i$ and $j$ are directly linked, then the corresponding element $A_{ij} = 1$. Otherwise, $A_{ij} = 0$. In addition, $A$, $F$ and $G$ can, in general, depend explicitly on time.

Two models are addressed in this work: a networked dynamical model to explain the development of excellent human performance \cite{DenHartigh2016} and a dynamical model of general intelligence \cite{VanderMaas2006}, both sharing great resemblance (Section \ref{section:interplay}).

\subsection{Model A: a networked dynamical model to explain the development of excellent human performance}
Ruud J.R.Den Hartigh et al. \cite{DenHartigh2016} were interested in the excellent level of performance of some individuals across different domains. They argued that the key to excellence does not reside in specific underlying components, but rather in the ongoing interactions among them and hence leading to the emergence of excellence out of the network integrated by genetic endowment, motivation, practice and coaching inter alia.

They attempted to render well-known characteristics of abilities leading to excellence: the absence of early indicators of ultimate exceptional abilities, the fact that a similar ability level may be shifted in time between individuals, the change of abilities during a person's life span and the existence of unique pathways leading to excellence, that is, individuals may have diverse ways to achieve it. 

They considered a networked dynamical model which can be mathematically defined as a set of coupled logistic growth equations, each of which represents the growth of a single variable, one of which being the domain-specific ability. The growth of the variable depends on the level already attained, available resources that remain relatively constant during development ($K_i$), resources that vary on the time scale of ability development, the degree in which a variable profits from the constant resources ($r_i$) and a general limiting factor ($C$): the ultimate carrying capacity, which captures the physical limits of growth. Moreover, $W_{ij}$ accounts for the effect of variable $x_i$ on $x_j$. 

\begin{figure}[!ht]
\centering

	\begin{subfigure}[t]{0.32\textwidth}
	\includegraphics[width=	1.1\linewidth]{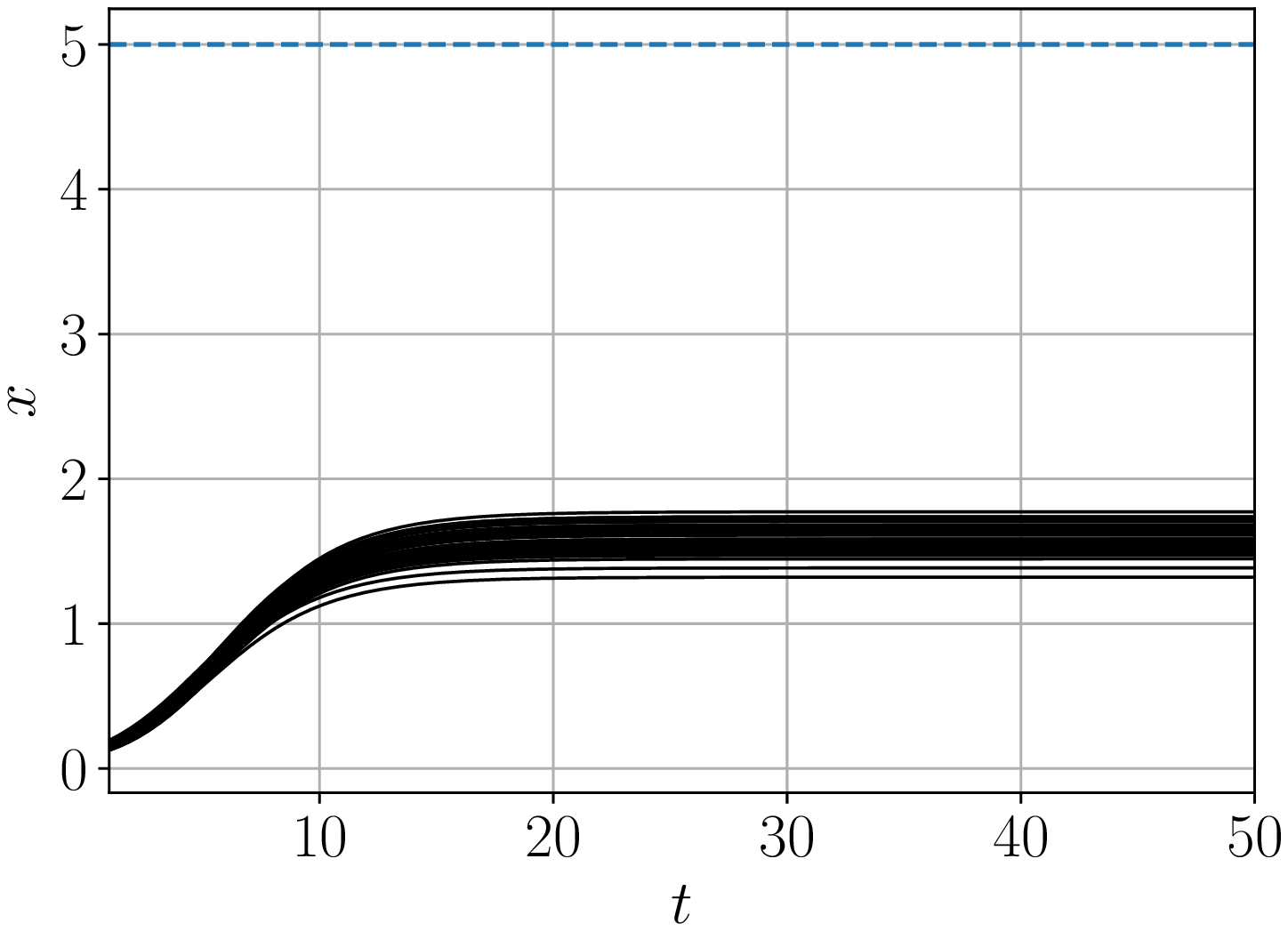}
	\label{metric_temporal}
	\end{subfigure}
	\hfill
	\begin{subfigure}[t]{0.32\textwidth}
	\includegraphics[width=	1.1\linewidth]{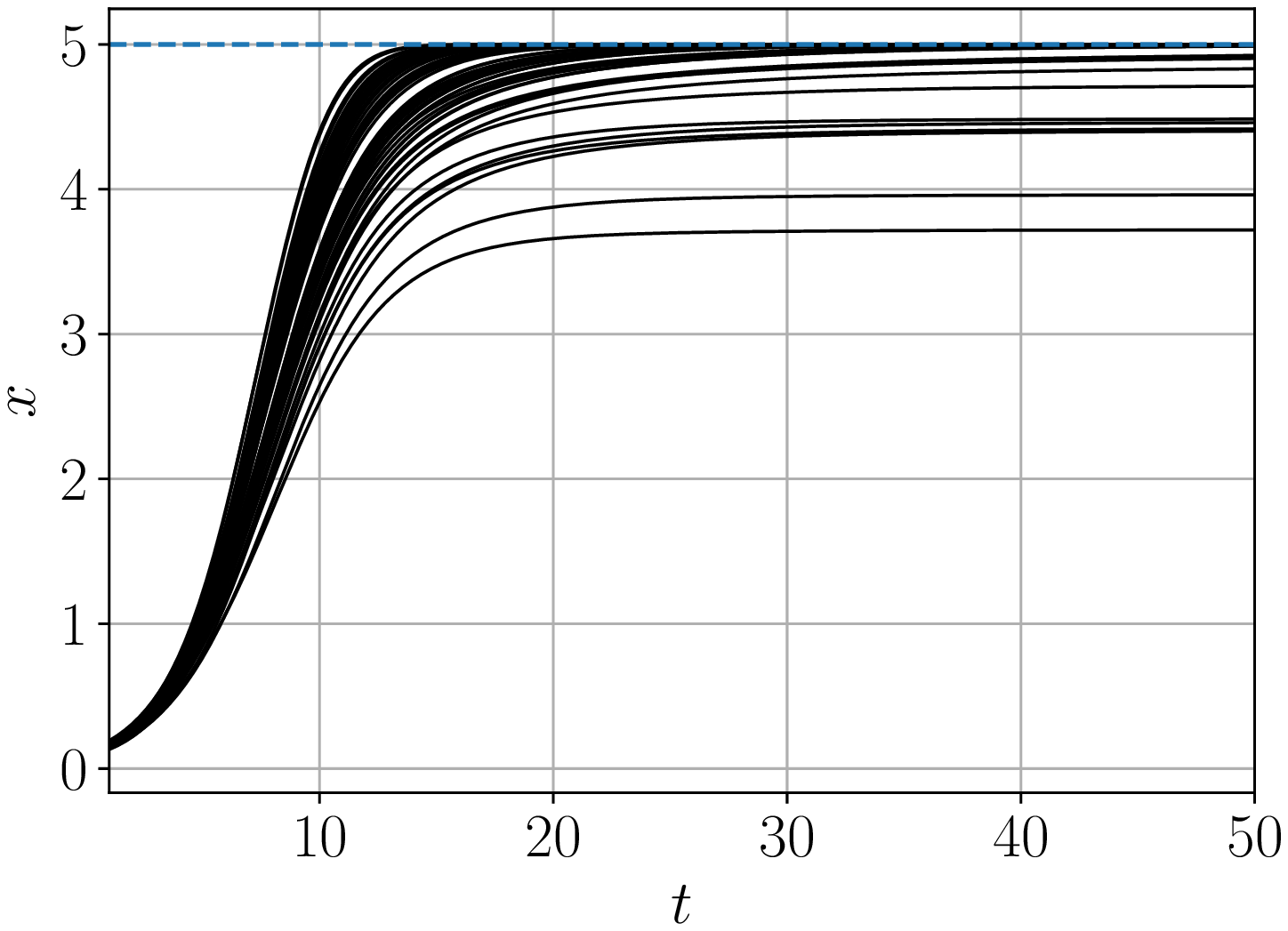}
	\label{mixed_temporal}
	\end{subfigure}
	\hfill
	\begin{subfigure}[t]{0.32\textwidth}
	\includegraphics[width=	1.1\linewidth]{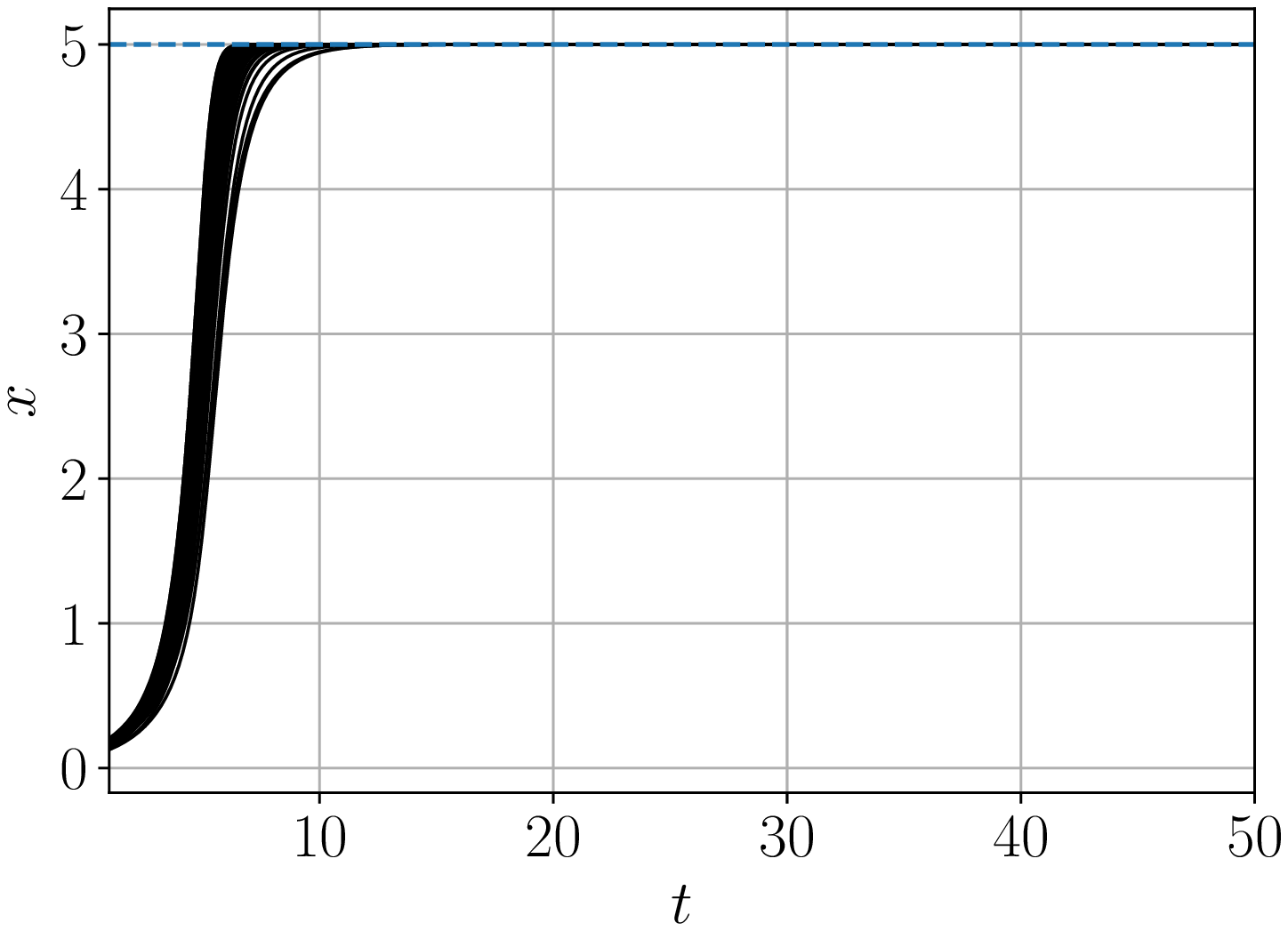}
	\label{optimal_temporal}
	\end{subfigure}
 \vspace{-0.5cm}
\caption{\small{Temporal evolution of all variables in the case of model A, described in Equation(\ref{modelA}), for increasing values of the links' weight: $w = 0.01$ (left), $w = 0.025$ (middle) and $w = 0.05$ (right). The system is attracted to one of the three possible global stable states, depending on both the parameters of the dynamic model and the topology: metric stable state, $\vec{x}(W_d)$ (left), mixed stable state (middle) and optimal stable state (right).  All plots are represented for an ER network with 50 nodes and edge probability $p=0.4$. Parameters are equally set to $C = 5$, $r = 0.5$, $K = 1.0$ for all nodes.}}
\label{temporal_evolution}
\end{figure}

Using (\ref{general}), model A can be written as:
\begin{equation}
\label{modelA}
\dot{x}_i = r_i x_i \left( 1 - \frac{x_i}{K_i}\right)\left(1 -\frac{x_i}{C}\right)+\sum_{j}W_{ji}x_ix_j \left(1 -\frac{x_i}{C}\right)
\end{equation}
Equation (\ref{modelA}) is better understood as a modified logistic growth:
\begin{equation}
\label{modelA_logistic}
\dot{x}_i = r_i x_i\left(1 -\frac{x_i}{C}\right) \left[ 1 - \frac{x_i - \displaystyle {K_i}/{r_i}\sum_{j}W_{ji}x_j}{K_i} \right]
\end{equation}
Figure \ref{temporal_evolution} shows 3 different possible temporal evolution of the system, determined by both the topology of the connection between variables and the parameters of the dynamical model.
\subsection{Model B: dynamical model of general intelligence}
Han L.J.van der Maas et al.\cite{VanderMaas2006} were concerned with the conceptualization and models of intelligence or cognitive abilities system by means of a general latent factor, as widely stated. They proposed an alternative explanation to the positive manifold based on a dynamical model built upon mutualistic interactions between cognitive processes, such as perception, memory, decision and reasoning, which are captured by psychometric tests scores to some extent. Such connections between items bring about another plausible explanation to the existence of one common factor, and thus need not correspond to an imposed latent process or actual quantitative variable, such as speed of processing or brain size.

Inspired by Lotka-Volterra models commonly used in population dynamics \cite{Hannan1977,Tsoularis2002}, they proposed to model the cognitive system as a developing ecosystem with primarily cooperative relations between cognitive processes. Variables ($x_i$) represent the distinct cognitive processes, which growth function is parametrized by the steepness of the growth ($r_i$) and the limited resources for each process ($K_i$). Matrix $W$ contains the relation between pair of processes, which they assume positive, i.e involved cognitive processes have mutual beneficial interactions.

Starting from uncorrelated initial conditions and parameters, that is, following uncorrelated random distributions, the dynamical connections between variables gradually lead the system to specific correlation patterns. 

Using (\ref{general}), model B can be written as:
\begin{equation}
\label{modelB}
\dot{x}_i = r_i x_i \left( 1 - \frac{x_i}{K_i}\right)+\sum_{j}W_{ji} \frac{r_i}{K_i}x_i x_j
\end{equation}
Equation (\ref{modelB}) is better understood as a modified logistic growth:
\begin{equation}
\label{modelB_logistic}
\dot{x}_i = r_i x_i \left[ 1 - \frac{x_i - \displaystyle \sum_{j}W_{ji}x_j}{K_i} \right]
\end{equation}
Section \ref{section:interplay} puts stress on the resemblance between (\ref{modelA_logistic}) and (\ref{modelB_logistic}).
\section{Interplay between dynamics and network topology}
\label{section:interplay}
A dynamical model which captures the network structure of the connection between variables using an expression similar to (\ref{general}) enables further analysis of the process considering the effect of topology, embodied in the adjacency matrix, $A$.

Neither of the two described models is geared toward a particular cognitive architecture or brain model with regard to connectivity structure. Rather, much effort is devoted to understanding the effect of non-zero correlations between the parameters of the models, or an heterogeneous landscape of parameters. The former approach, however, requires certain constrains which, in general, may not be easy to proof. Alternatively, we consider the dynamical model to be parametrized by an homogeneous configuration, i.e. all nodes with equally fixed parameters, and explore the role of different connectivity structures between variables, which can be mapped on a network.

Although a dynamical model describes the temporal evolution of several variables, we are usually concerned with the final state.
  
\subsection{Mapping between models through weights rescaling}
\label{stability_analysis}
The space of parameters is large and hence so is the number of possible stable states. However, we focus our interest on solutions given by one unique analytical expression. Therefore, depending on the stability conditions (Section \ref{stability_conditions}) we can distinguish two of such stable states. For model A, an optimal stable solution, $\vec{x}(C)$, is achieved when all variables reach the maximum allowed value:
\begin{equation}
\label{modelA_fixed_C}
x(C)_i \equiv C \ \forall i
\end{equation} 
where we assume $C<k_i$. 

Otherwise, final state, $\vec{x}(W_d)$, is determined by  matrix expression (\ref{modelA_fixed_K}):
\begin{equation}
\label{modelA_fixed_K}
\vec{x}(W_d) = (\mathbb{I}-\mathbb{W}_d)^{-1}\vec{K}
\end{equation} 
$\mathbb{W}_d$ matrix in expression (\ref{modelA_fixed_K}) captures the entanglement between network topology, $W$, and the parameters of the dynamical model. The influence of variable $i$ on variable $j$ is thus rescaled by its carrying capacity, $K_i$ and growing rate, $r_i$, as follows:
\begin{equation}
\label{W_d}
[\mathbb{W}_d]_{ij} \equiv \frac{K_i}{r_i}W_{ji}
\end{equation}
All intermediate states, which lay in the transition between metric and optimal stable states, are called mixed stable state. 

Analogously, for model B, there is one unique stable state, $\vec{x}(W)$:
\begin{equation}
\label{modelB_fixed_K}
\vec{x}(W) = (\mathbb{I}-W^T)^{-1}\vec{K}
\end{equation} 
Expressions (\ref{modelA_fixed_K}) and (\ref{modelB_fixed_K}), referred to stable states, for model A and B, respectively, are equivalent under rescaling (\ref{W_d}). Moreover, we highlight the absence of initial conditions in the attractor state.

The case when parameters are constant throughout variables, i.e, when $K_i \equiv K \ \forall i$, $r_i \equiv r \ \forall i$ and $W_{ij} \equiv w \ \forall (i,j)$, may enable an explicit average solution to  (\ref{modelB_fixed_K}). An individual can be characterized by the average of the achived values of all variables. We follow the notation:
\begin{equation}
\bar{x}^* \equiv \frac{1}{N}\sum^{N}_{i} x^*_i
\label{average}
\end{equation}

For a complete network: 
\begin{eqnarray}
\label{modelB_complete}
\bar{x}^*  = \frac{K}{1-w(N-1)} & & Var \left[ x(W)_i\right] = 0 \ \forall i
\end{eqnarray}
For an Erd\"{o}s-R\'{e}nyi network (Appendix \ref{appendix:solution_ER}):
\begin{eqnarray}
\label{modelB_ER}
\bar{x}^*  = \frac{K}{1-w \left\langle k \right\rangle} & & Var \left[ x(W)_i\right] \approx K^2w^2 Var \left[ k\right] + O(w^3) \ \forall i
\end{eqnarray}
where $Var \left[ k\right]$ is the degree of nodes variance.

Average solutions to (\ref{modelA_fixed_K}) for a complete network and an Erd\"{o}s-R\'{e}nyi network are equivalent to (\ref{modelB_complete}) and (\ref{modelB_ER}), respectively, with $w_d \equiv \displaystyle\frac{K}{r}w$.
\subsection{Katz-Bonacich centrality as stable state}
\label{katz_centrality}
Centrality measures seek the most important or central nodes in a network \cite{Newman2010}. Among many possible centralities, the generalized Katz-Bonacich centrality \cite{Katz1953} is given by the solution of:
\begin{equation}
\label{katz_definition}
x_i = \alpha \sum_j A_{ji}x_j + \beta_i
\end{equation}  
Solving (\ref{katz_definition}), the vector $\vec{x}$ of centralities is given by:
\begin{equation}
\label{katz_matrix}
\vec{x} = (\mathbb{I}-\alpha A^T)^{-1}\vec{\beta}
\end{equation}
Unlike eigenvector centrality, Katz-Bonacich centrality solves the issue of zero centrality values for acyclic or not strongly-connected networks by introducing a constant term $\beta_i$ for each node. Therefore, Katz-Bonacich centrality gives each node a score proportional to the sum of the scores of its neighbours plus a constant value. $\alpha$ parameter rules the balance between the first term in (\ref{katz_definition}), which is the normal eigenvector centrality \cite{Newman2006}, and the second. 
The longest walks become more significant as $\alpha$ increases and hence the global topology of the network is considered, resembling eigenvector centrality. On the contrary, small values of $\alpha$ make Katz-Bonacich centrality a local measure which approaches degree centrality. When $\alpha \to 0$, $\vec{x}=\vec{\beta}$ and as $\alpha$ increases so do the centralities until they diverge when:
\begin{equation}
\label{katz_divergence}
\alpha = \frac{1}{\lambda(A)_{\max}}
\end{equation}
where $\lambda(A)_{\max}$ is the maximum eigenvalue of $A$ matrix. Hence, Katz-Bonacich centrality is defined as long as $\alpha < \lambda(A)_{\max}^{-1}$ \cite{Newman2010}.

Equivalently, for weighted networks, generalized Katz-Bonacich centrality is defined as:
\begin{equation}
\label{katz_matrix_W}
\vec{x} = (\mathbb{I}-\alpha W^T)^{-1}\vec{\beta}
\end{equation}
and $\alpha < \lambda(W)_{\max}^{-1}$.

Equation (\ref{katz_matrix_W}) can also be expanded to:
\begin{equation}
\label{katz_walks}
\vec{x} = (\mathbb{I}+\alpha W^T + \alpha^2 (W^T)^2+ \alpha^3 (W^T)^3 + \cdots)\vec{\beta} = \sum_{p=0}^{p=\infty}(\alpha W^T)^p \vec{\beta}
\end{equation}
Element $[(W^T)^p]_{ij}$ in (\ref{katz_walks}) stands for the number of walks of length $p$ from node $j$ to node $i$ taking the strength of connections into account. This value is attenuated by a factor $\alpha^p$ and hence $[\sum_{p=0}^{p=\infty}(\alpha W^T)^p]_{ij}$ accounts for the strength of all walks from node $j$ to node $i$, with greater weakening as $p$ gets larger.

Comparing (\ref{katz_matrix_W}) with (\ref{modelA_fixed_K}) or (\ref{modelB_fixed_K}) we conclude that generalized Katz-Bonacich centrality vector is the exact solution of the stable state of model B with $\alpha \equiv 1$ and $\vec{\beta} \equiv \vec{K}$. Furthermore, when rescaling (\ref{W_d}) is considered, so it is of model A or any other model which dynamics can be included in the weighted adjacency matrix in a similar way.

Therefore, variables which score greater according to generalized Katz-Bonacich centrality, achieve optimal values on the long run. For this reason we call stable state (\ref{modelA_fixed_K}) as \textit{``metric" stable state}.

For further discussion, we recall that a subset of centrality measures can also be interpreted as the stable state of a random walk along a network. Namely, generalized Katz-Bonacich centrality is the stable state of a biased random walk on a network for non-conservative processes \cite{Ghosh2014,Borgatti2005}.

\subsection{Stability conditions}
\label{stability_conditions}
For model A, stability analysis is rather complex since many different stable states may exist, depending on a sizeable number of parameters which characterize both the topology and the dynamics. Nevertheless, we focus our interest on the most extreme situations: the optimal stable state, $\vec{x}(C)$, given by (\ref{modelA_fixed_C}) and the metric stable state, $\vec{x}(\mathbb{W}_d)$, given by (\ref{modelA_fixed_K}). All other configurations are described by a mixed pattern which falls between optimal and metric stable states.

$\vec{x}(C)$ solution is stable when (See Appendix \ref{appendix:stability}):
\begin{equation}
r_i\left(1-\frac{C}{K_i}\right)+C\sum_jW_{ji}>0 \ \forall i\label{stability_optimal}
\end{equation}
Or, using rescaled weighted matrix $\mathbb{W}_d$ defined in (\ref{W_d}), when:
\begin{equation}
\sum_j [\mathbb{W}_d]_{ji} > 1-\frac{K_i}{C} \ \forall i
 \label{stability_optimal_rescaling}
\end{equation}
Provided that a given node $i$ does not meet condition (\ref{stability_optimal_rescaling}), stable state is no longer $\vec{x}(C)$ and hence, starting from node $i$, nodes will start getting values lower than the $C$ threshold.

On the other hand, $\vec{x}(\mathbb{W}_d)$ solution is stable when (Appendix \ref{appendix:stability}):
\begin{equation}
\lambda_{\max}(\mathcal{S})<0
\label{stability_metric}
\end{equation}
where $\lambda_{\max}(\mathcal{S})$ is the maximum eigenvalue of matrix $\mathcal{S}$, which is defined as follows:
\begin{equation}
\label{S_matrix}
\mathcal{S} \equiv \mathbb{D}(\mathbb{I}-\mathbb{W}_d) \ ; \ \mathbb{D}_{ij} \equiv -x(\mathbb{W}_d)_i\left( 1 - \frac{x(\mathbb{W}_d)_i}{C}\right)\delta_{ij}
\end{equation}
Eigenvalues of $\mathcal{S}$ depend explicitly on $\vec{x}(\mathbb{W}_d)$ and therefore can only be computed numerically (See Equation \ref{metric_stability}). However, Perron-Frobenius theorem \cite{UnnikrishnaPillai2005} allows us to obtain an upper threshold for $\lambda_{\max}(\mathcal{S})$ analytically:
\begin{equation}
\label{perron-frobenius}
\lambda_{\max}(\mathcal{S})<\max \left[\sum_j \mathcal{S}_{ij}\right]
\end{equation}
Equations (\ref{stability_metric}) and (\ref{perron-frobenius}) imply the existence of an upper bound for the stability condition of metric stable state (Appendix \ref{appendix:stability}):
\begin{equation}
\left(\sum_j [\mathbb{W}_d]_{ij}\right)_{\max} < 1 
 \label{stability_upper_threshold}
\end{equation}
For model B, stability analysis concerns only the metric stable state (\ref{modelB_fixed_K}) and stability conditions are given by (\ref{stability_metric}) with:
\begin{equation}
\label{S_matrix_modelB}
\mathcal{S} \equiv \mathbb{D}(\mathbb{I}-W^T) \ ; \ \mathbb{D}_{ij} \equiv -\frac{r_i}{K_i} x(W)_i\delta_{ij}
\end{equation}

\begin{figure}[!ht]
\centering
	\begin{subfigure}[t]{0.48\textwidth}
	\includegraphics[width=	1.1\linewidth]{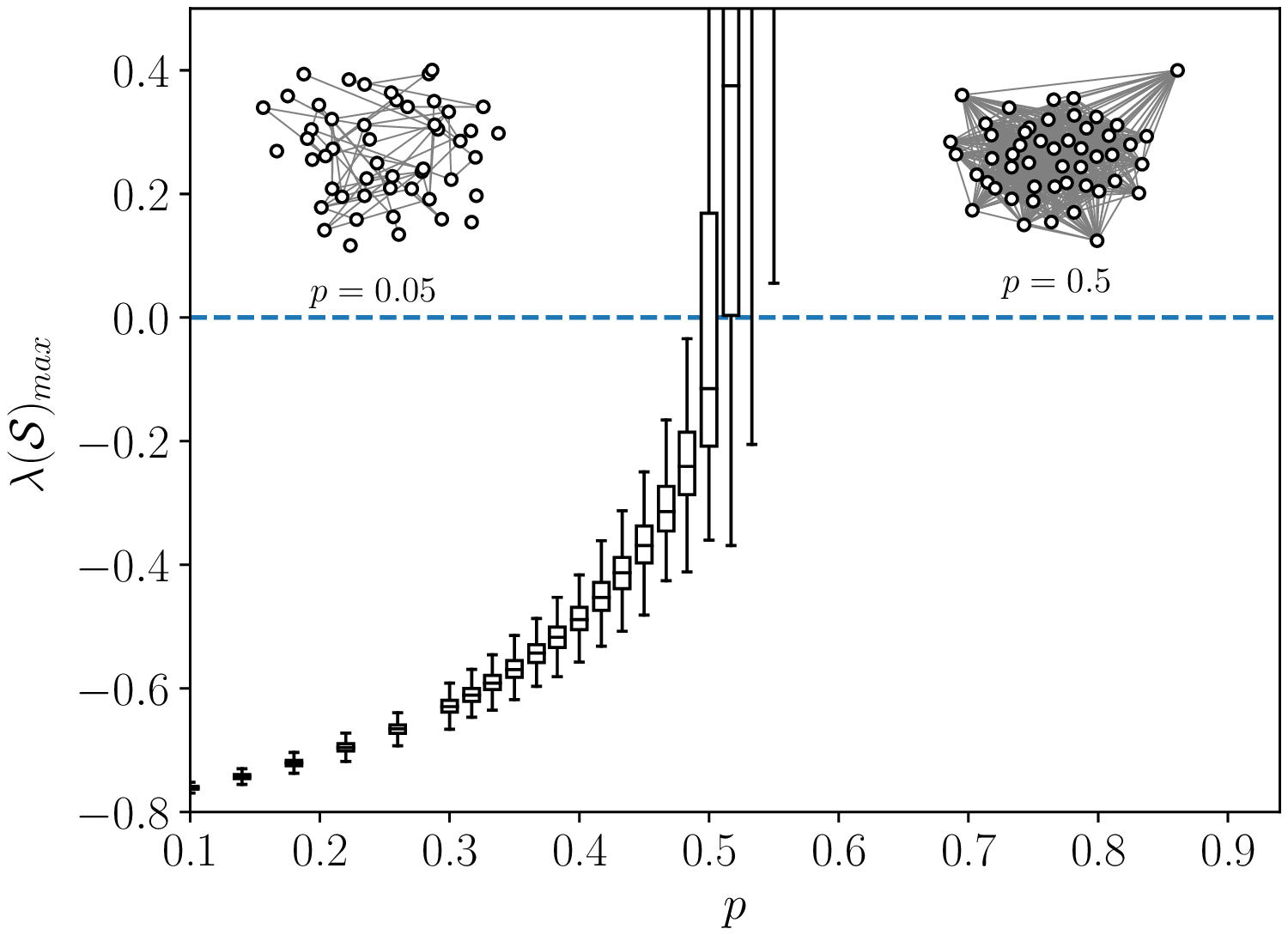}
	\label{stability_ER}
	\end{subfigure}
	\hfill
	\begin{subfigure}[t]{0.48\textwidth}
	\includegraphics[width=	1.1\linewidth]{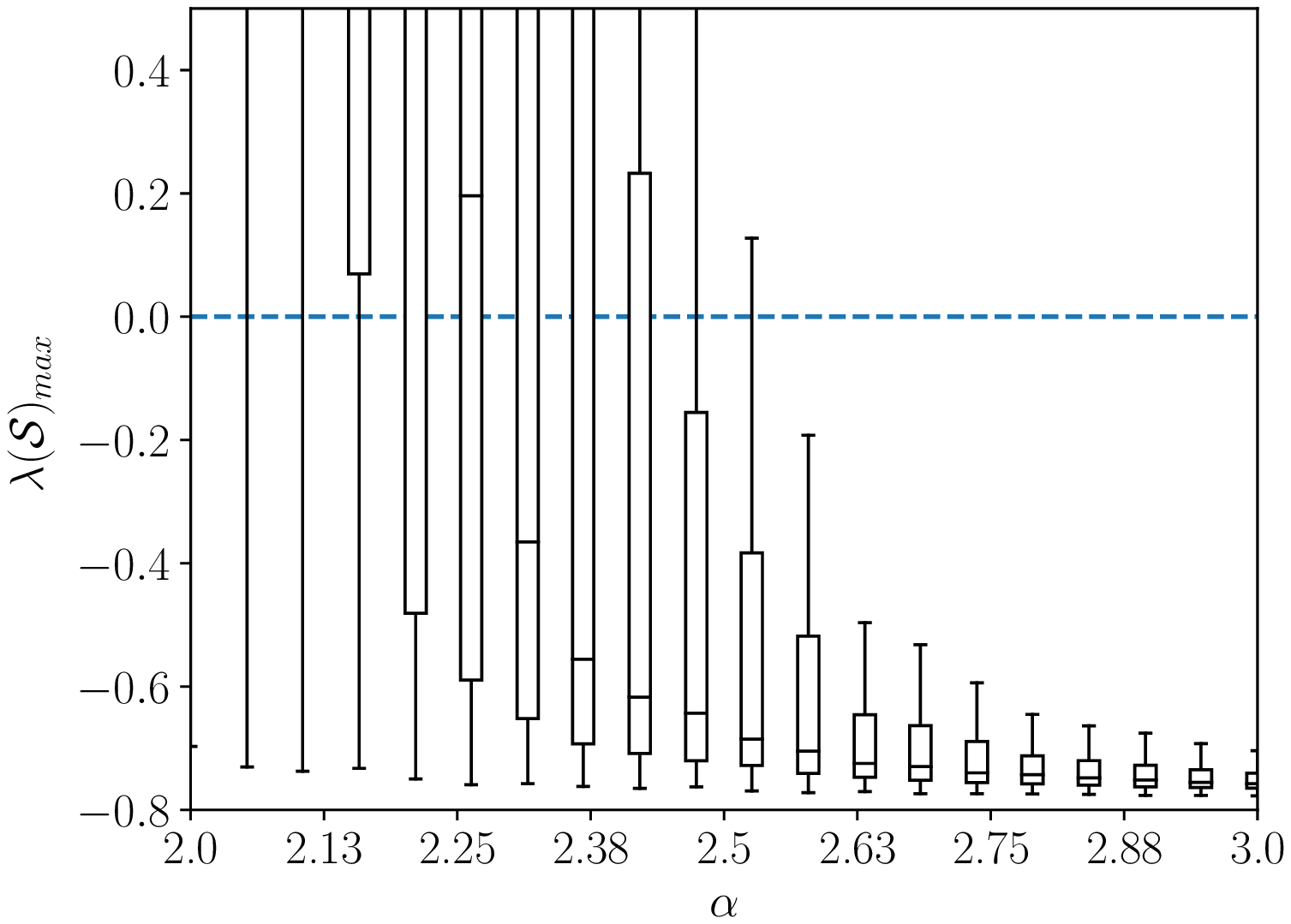}
	\label{stability_heterogeneous}
	\end{subfigure}
	\hfill
	\begin{subfigure}[t]{0.48\textwidth}
	\includegraphics[width=	1.1\linewidth]{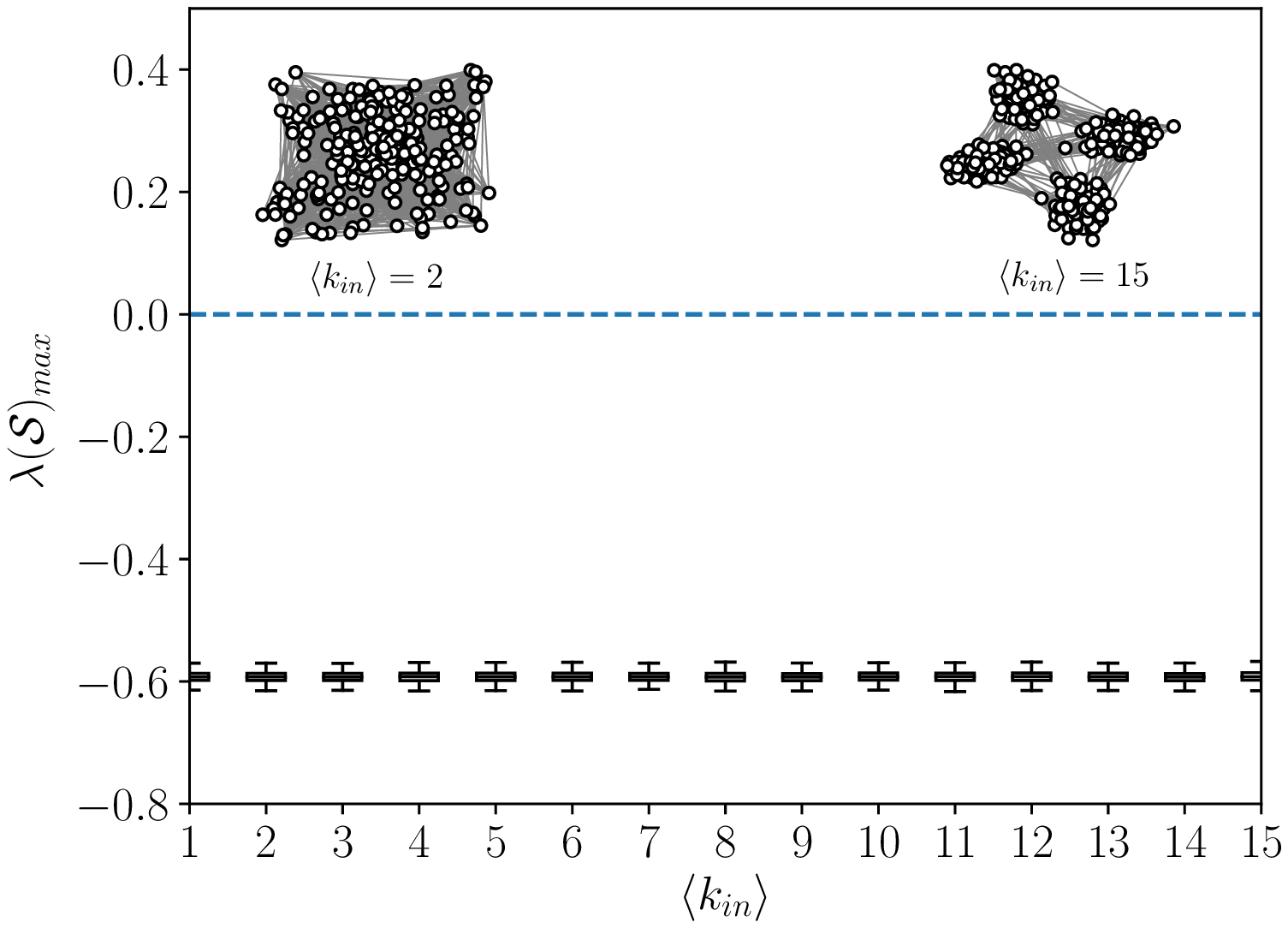}
	\label{stability_newman}
	\end{subfigure}
\vspace{-0.5cm}
\caption{\small{(Color online) Set of boxplots of $\lambda(\mathcal{S})_{\max}$, defined in (\ref{S_matrix}), for different values of the characteristic parameter of the network: edge probability $p$ for an Erd\"{o}s-R\'{e}nyi network of size $N=50$ (upper left), degree distribution exponent $\alpha$ for an heteogeneous network of size $N=200$ (upper right) and intra-module degree $\left\langle k_{in} \right\rangle$ for a Newman modular network of size $N=256$ and $\left\langle k_{total} \right\rangle =16$ (bottom). Metric state is given by (\ref{modelA_fixed_K}) and it is stable only if $\lambda(\mathcal{S})_{\max}<0$ (dashed blue line). For each value of $\{\mu \}$ we make 1000 realization of the corresponding network class, each one representing one single individual, to capture the distribution of the maximum eigenvalue of matrix $\mathcal{S}$. Edge probability, $p$, of an Erd\"{o}s-R\'{e}nyi network makes for a narrower threshold of stability than $\alpha$ exponent of an heterogeneous network. Conversely, $\left\langle k_{in} \right\rangle$ parameter of Newman modular network does not lead to any change in stability. Parameters of the model are set for all nodes to $C=5$ and $K=1$ for all networks, $r=1$ and $w=0.03$ for Erd\"{o}s-R\'{e}nyi network, $r=0.5$ and $w=0.025$ for heterogeneous network and $r=0.05$ and $w=0.015$ for Newman modular network.}}
\label{fig:metric_stability}
\end{figure}
From Figure \ref{fig:metric_stability}, we conclude that network topology is enough to obtain information about the stability of the system. Due to the fact that we are in a situation of homogeneous configuration, $\lambda (\mathcal{S})_{\max}$ is tightly connected to $\lambda (W)_{\max}$, although it is not the same as seen in (\ref{S_matrix}) and (\ref{S_matrix_modelB}). Concerning Erd\"{o}s-R\'{e}nyi network (Figure \ref{fig:metric_stability} (upper left)), there is little variability in $\lambda (\mathcal{S})_{\max}$ and therefore the critical value of the $p$ parameter for which stability changes, $p_{C}$, is confined within a narrow range. Using (\ref{stability_upper_threshold}) and assuming a Poisson degree distribution we can obtain an approximate value for $p_C$ in case of a homogeneous configuration:
\begin{equation}
\left(\sum_j [\mathbb{W}_d]_{ij}\right)_{\max} \approx w \left[ \left\langle k \right\rangle + \sqrt{Var(k)} \right] \approx w  \left[ pN + \sqrt{pN} \right] < 1 \Rightarrow p < p_C
\label{threshold_approx}
\end{equation}
where $p_C \equiv \displaystyle \frac{2N/w + N - N\sqrt{1+4/w}}{2N^2}$.
\vspace{0.2cm}

The critical value of $p$ obtained from (\ref{threshold_approx}) when the parameters are the same as for the Erd\"{o}s-R\'{e}nyi network of Figure \ref{fig:metric_stability} gives $p_C \approx 0.56$, which is in agreement with the numerical solution.

Conversely, stability with regard to heterogeneous network is rather diffuse (Figure \ref{fig:metric_stability} (upper right)). The broad spectrum of $\lambda (W)_{\max}$ \cite{Goh2001a} is somehow captured by the variability on $\lambda (\mathcal{S})_{\max}$. Consequently, there exist outlier networks which eventually break stability condition. Nevertheless, they become less frequent as $\alpha$ exponent increases. The effect of $\left\langle k_{in} \right\rangle$ in a Newman modular network is completely different (Figure \ref{fig:metric_stability} (below)): in spite of being a parameter which rules the modularity of the network, the average total degree, $\left\langle k_{total} \right\rangle = \left\langle k_{in} \right\rangle + \left\langle k_{out} \right\rangle$, is still fixed and hence, both $\mathcal{S}$ matrix and solutions  (\ref{modelA_fixed_K}) and (\ref{modelB_fixed_K}) remain essentially the same, except fluctuations.
 
\section{Unveiling correlation structures from network topology}
\label{section:correlations}
Provided that condition (\ref{stability_metric}) is met, the stable state is given by (\ref{modelA_fixed_K}) and (\ref{modelB_fixed_K}), for model A and B, respectively. This latter result provides one individual with as many values as existing variables. However, studies based on large batteries of psychometric tests rely on a sample from a population, made up of tens to thousands of individuals, from where inter-variables correlations are inferred. Section \ref{generating_correlations} describes the generation of distinct individuals from the models and handling of the correlation matrix out of them.  
\subsection{From dynamics to correlation matrix}
\label{generating_correlations}

These models have been studied by assuming that parameters corresponding to different variables are correlated. If, in addition, all variables are considered to be equally interconnected in a mutualistic scenario, that is, positive interactions and a complete network topology is considered, it is possible to recover well-known correlation structures\cite{VanderMaas2006}. Alternatively, our hypothesis is that the network topology of variables is sufficient to recover equivalent results concerning correlations and also enables us to avoid stronger constraints on parameters.

Since we are interested in studying the effect of topology, each individual is considered to be one random instance of the same network class. In other words, a new individual is obtained out of the possible random $G(N,\{\mu\})$ and the corresponding correlation matrix is characterized by fixed values of $N$ and $\{\mu\}$, where $N$ is the number of variables or nodes and $\{\mu\}$ is the set of characteristic attributes of a given network model. For instance, a network generated by Erd\"{o}s-R\'{e}nyi model has only one parameter, i.e, $\{\mu\}=p$ (Section \ref{section:network}). 

Despite each variable holding its own set of parameters, we make all variables equivalent in what we call \textit{homogeneous configuration}, such that individual differences result only from topological properties:
\begin{eqnarray}
K_i \equiv K , \ r_i \equiv r & \forall i \ \ \ W_{ij} \equiv w & \forall (i,j) 
\label{homogeneous}
\end{eqnarray} 

Independently of others, each individual reaches a stable state given by (\ref{modelA_fixed_K}) or (\ref{modelB_fixed_K}), for model A or B, respectively, as long as condition (\ref{stability_metric}) is true. If this is not the case (See Section \ref{stability_conditions}), model A requires full temporal evolution being simulated and model B comes to divergence and hence the considered configuration is not acceptable.
\tikzstyle{decision} = [diamond, draw, fill=blue!20, 
    text width=4.5em, text badly centered, node distance=3cm, inner sep=0pt]
\tikzstyle{block} = [rectangle, draw, fill=blue!20, 
    text width=7em, text centered, rounded corners, minimum height=4em]
\tikzstyle{line} = [draw, -latex']
\tikzstyle{cloud} = [draw, ellipse,fill=red!20, node distance=3cm,
    minimum height=2em]

\begin{figure}[h]
\centering
\scalebox{0.7}{
\begin{tikzpicture}[node distance = 2cm, auto]
    \node [block] (network) {Generate network $G(N,\{\mu\})$};
    \node [block, below of=network] (parameters) {Initialize parameters: homogenous};
    \node [decision, below of=parameters] (decide) {Is metric solution stable?};
    \node [block, below left of=decide, node distance=3.5cm] (run) {Numerical solution (temporal evolution)};
    \node [block, below right of=decide, node distance=3.5cm] (analytic) {Analytic solution (Equation \ref{modelA_fixed_K}) or (\ref{modelB_fixed_K})};
    \node [block, below of=decide, node distance=5cm] (solution) {Get stable state};
    \node [decision, below of=solution] (population) {Any left individual from population?};
    \node [block, below of=population, node distance=3cm] (correlation) {Get correlation matrix};
     
     \path [line] (network) -- (parameters);
     \path [line] (parameters) -- (decide);
     \path [line] (decide) -- node[midway,above, sloped] {no} (run);
     \path [line] (decide) -- node[midway,above, sloped] {yes} (analytic);
     \path [line] (run) -- (solution);
     \path [line] (analytic) -- (solution);
     \path [line] (solution) -- (population);
     \path [line] (population) -- node[midway,above, sloped] {no} (correlation);
     \path [line] (population)  edge   [bend left=110]node[midway,above, sloped] {yes} (network); 
     
\end{tikzpicture}
}
\caption{Flowchart to generate correlation matrix of variables out of a population of individuals}
\label{fig:flowchart}
\end{figure}
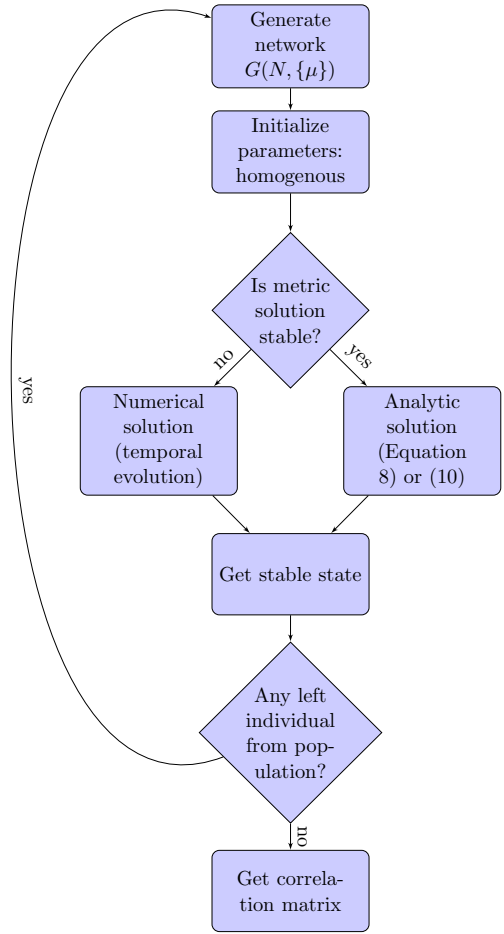
Once stable states of the entire population are obtained, the computation of pairwise correlations between variables is straightforward. In line with most studies conveyed by psychometrics, Pearson standard correlation coefficient is used to finally get the correlation matrix \cite{Cohen2009}, although other methods may excel this procedure \cite{Spearman1904a,DeWinter2016}.

In conclusion, the variability within one same family of networks $\{G(N,\{\mu\})\}$ brings about individual differences concerning variables which entail particular patterns or structures of the correlation matrix (\ref{correlation_function}), with no explicit constrain on parameters, but considering them as homogeneous both between variables and individuals (See Figure \ref{fig:flowchart}). 
\begin{equation}
\label{correlation_function}
\mathit{R(N,\{\mu\})} = f \left[ \{G(N,\{\mu\}) \}\right]
\end{equation}

\begin{figure}[!ht]
\centering
	\begin{subfigure}[t]{0.48\textwidth}
	\includegraphics[width=	1.1\linewidth]{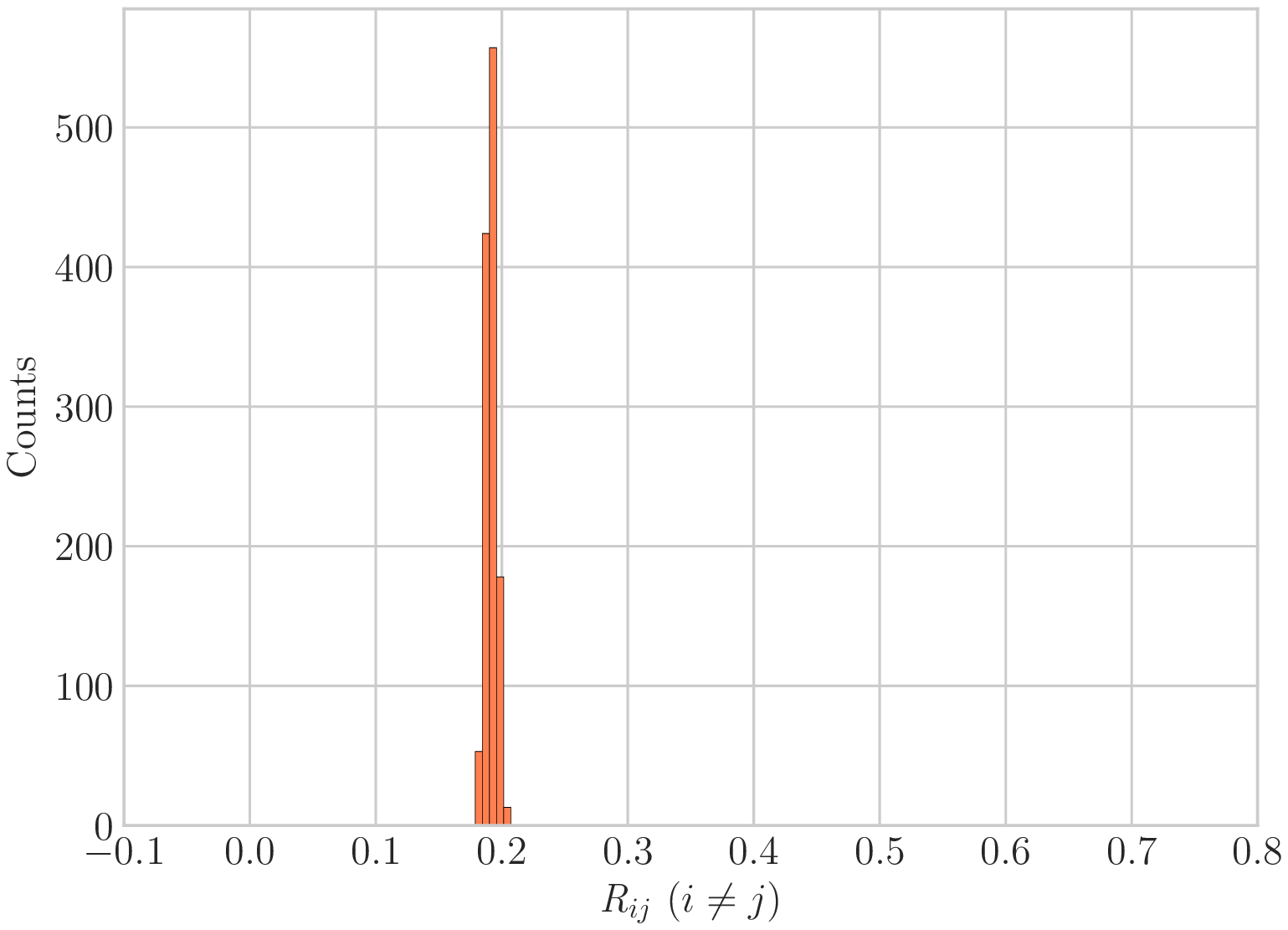}
	\label{correlations_ER}
	\end{subfigure}
	\hfill
	\begin{subfigure}[t]{0.48\textwidth}
	\includegraphics[width=	1.1\linewidth]{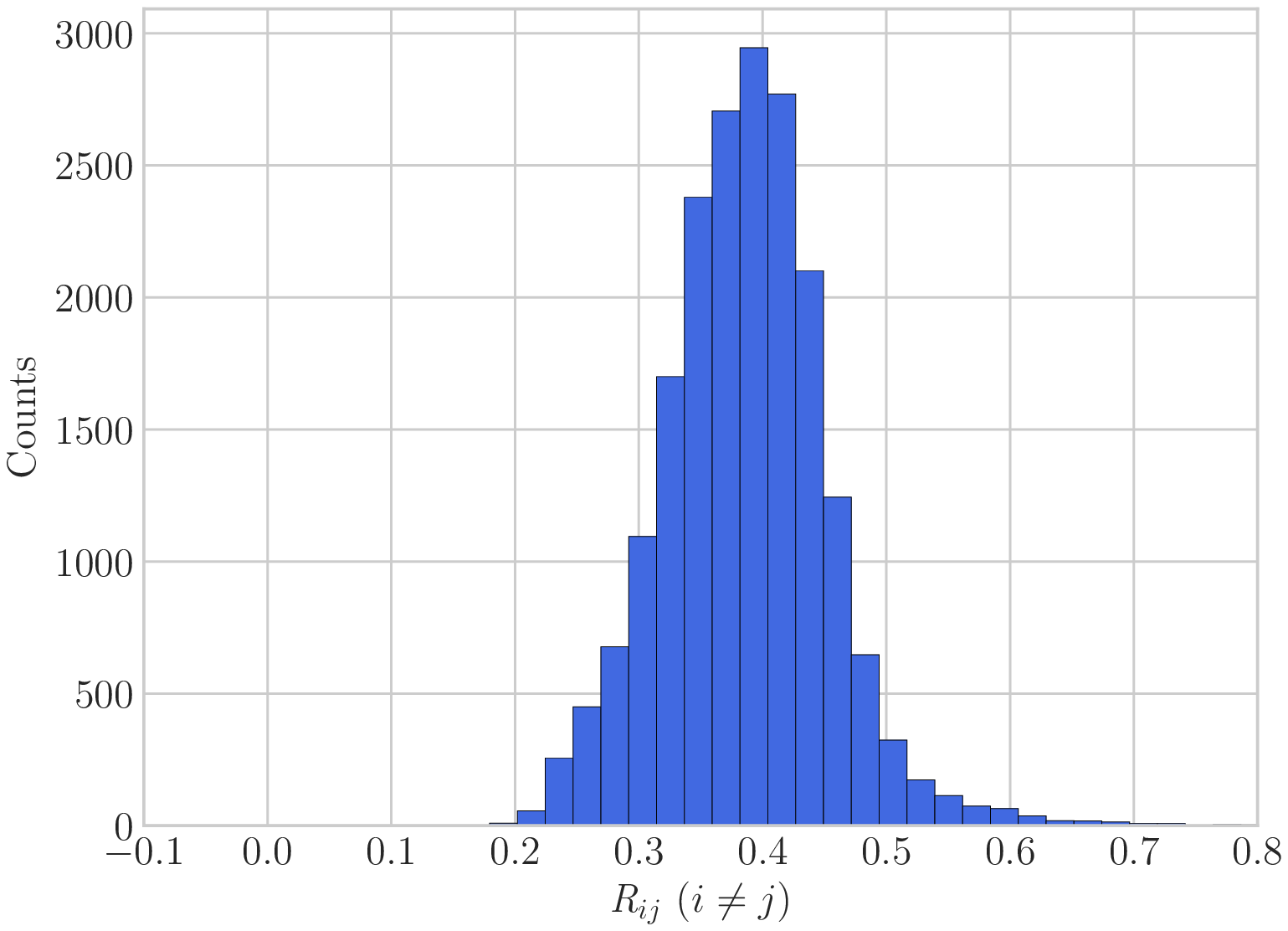}
	\label{correlations_heterogeneous}
	\end{subfigure}
	\begin{subfigure}[t]{0.48\textwidth}
	\includegraphics[width=	1.1\linewidth]{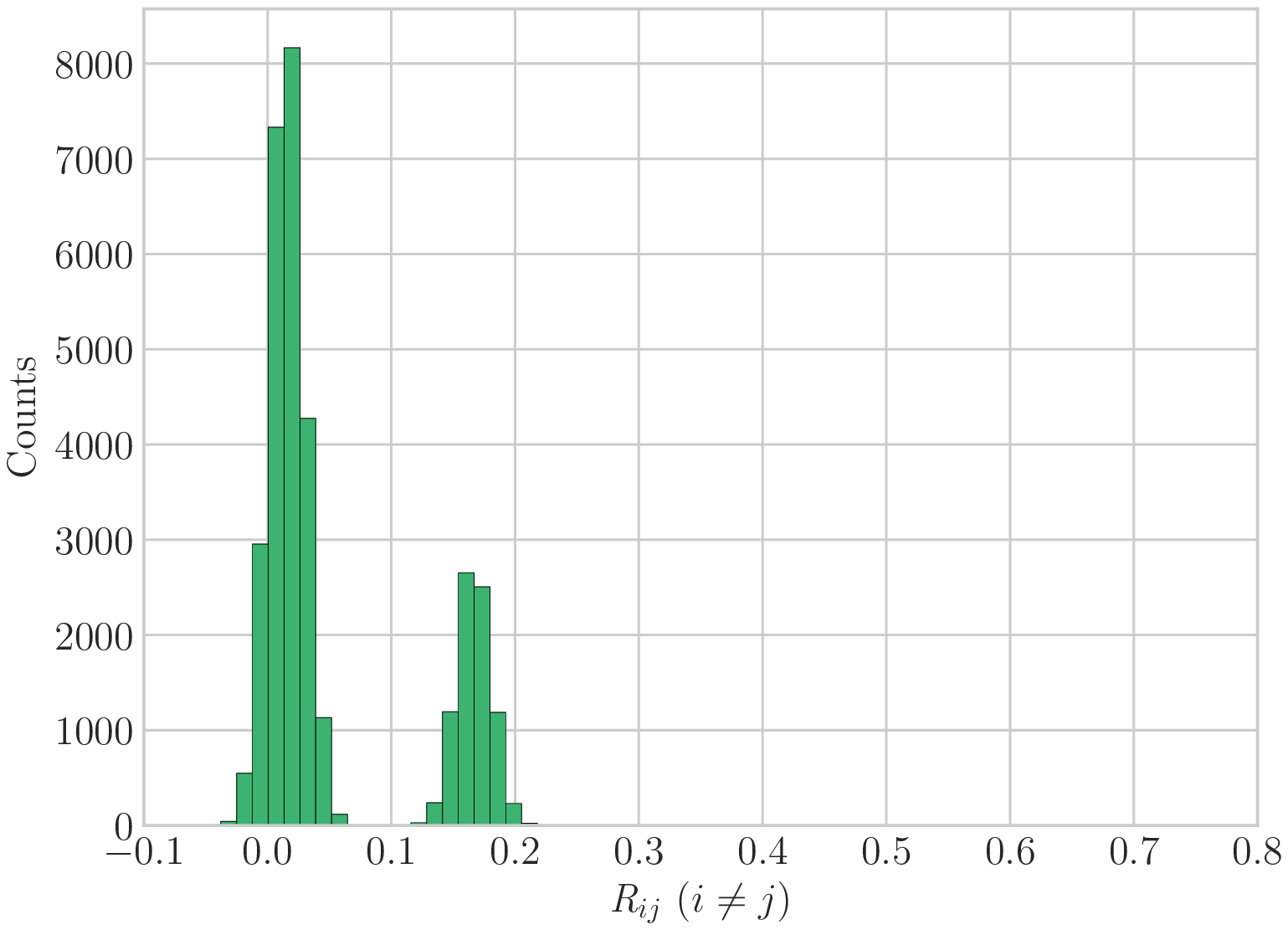}
	\label{correlations_newman}
	\end{subfigure}
	\vspace{-0.5cm}
\caption{\small{Histogram of the correlation between variables, $\mathit{R}_{ij}$, for 3 distinct network topologies: Erd\"{o}s-R\'{e}nyi of size $N=50$ and edge probability $p=0.4$ (upper left), heterogeneous network of size $N=200$ and degree distribution exponent $\alpha=2.9$ (upper right) and Newman modular of size $N=256$, $\left\langle k_{in} \right\rangle=15$ and $\left\langle k_{total} \right\rangle=16$ (bottom). The histogram captures the effect of $\{G(N,\{\mu\}) \}$ on $\mathit{R(N,\{\mu\})}$. Correlation matrix is computed from analytic metric stable state (\ref{modelA_fixed_K}) of 5000 generated individuals. The distribution of correlations between variables connected as an ER network is peaked at a certain value, whereas assuming an heterogeneous network leads to a much broader spectrum of correlations values. Newman modular network brings about two clear peaks, corresponding to intracluster and intercluster correlations. Parameters of the model are set for all nodes to $C=5$ and $K=1$ for all networks, $r=1$ and $w=0.03$ for Erd\"{o}s-R\'{e}nyi network, $r=0.5$ and $w=0.025$ for heterogeneous network and $r=0.05$ and $w=0.015$ for Newman modular network.}}
\label{correlations_histogram}
\end{figure}

From figure \ref{correlations_histogram} we conclude that the positive manifold, i.e positive correlations, $\mathit{R}_{ij} > 0$, can come out regardless of the topology, as long as $\{\mu\}$ of the network is properly set. Thus, connectivity by itself allows positive interactions, regardless of the structure and independently of further constrains on dynamic parameters. However, the structure of the correlation matrix indeed relies on the topology: Erd\"{o}s-R\'{e}nyi network's correlations histogram  displays a narrow symmetric peak which captures the homogeneity of the topology, i.e all nodes being equivalent. On the contrary, heterogeneous network's correlations histogram follows a much wider asymmetric distribution, with a longer tail for larger values. This behaviour captures the presence of few hubs, which lead to non-trivial correlations structures, as we will see in Section \ref{statistical_models}. Finally, Newman modular network  presents two characteristic peaks corresponding to intracluster, i.e within the same cluster, and intercluster, i.e between different clusters, correlations. As $\left\langle k_{in} \right\rangle$ increases, both peaks become undistinguishable and pattern more closely resembles an Erd\"{o}s-R\'{e}nyi network. For large $\left\langle k_{in} \right\rangle$, intercluster correlations tend to $0$ value.

So far, in addition to considering an homogeneous configuration of parameters, we have constrained ourselves to solutions within the regime where metric stable solution is stable. Nevertheless, we expect richer phenomena when neither of the two restrictions exists. If we allow $K$ parameter to be a random variable distributed among variables following a non-correlated normal distribution, $N(\mu_K, \sigma_K)$, increasing values of $\sigma_K$ lead to lower values of $\left\langle \mathit{R}_{ij} \right\rangle$, the average correlation between variables.

\begin{figure}[h!]
\centering
	\begin{subfigure}[t]{0.48\textwidth}
	\includegraphics[width=	1.1\linewidth]{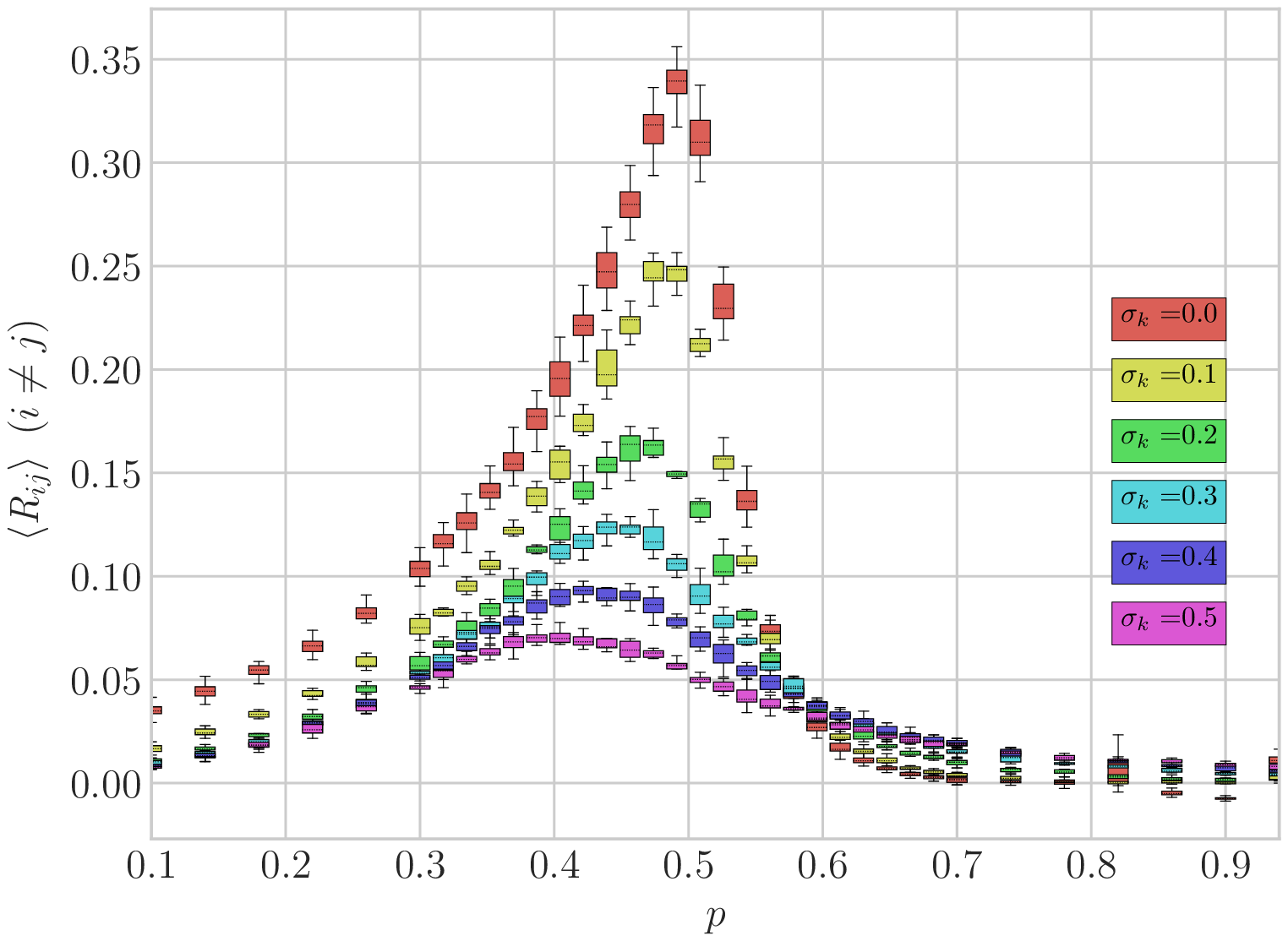}
	\label{R_ER}
	\end{subfigure}
	\hfill
	\begin{subfigure}[t]{0.48\textwidth}
	\includegraphics[width=	1.1\linewidth]{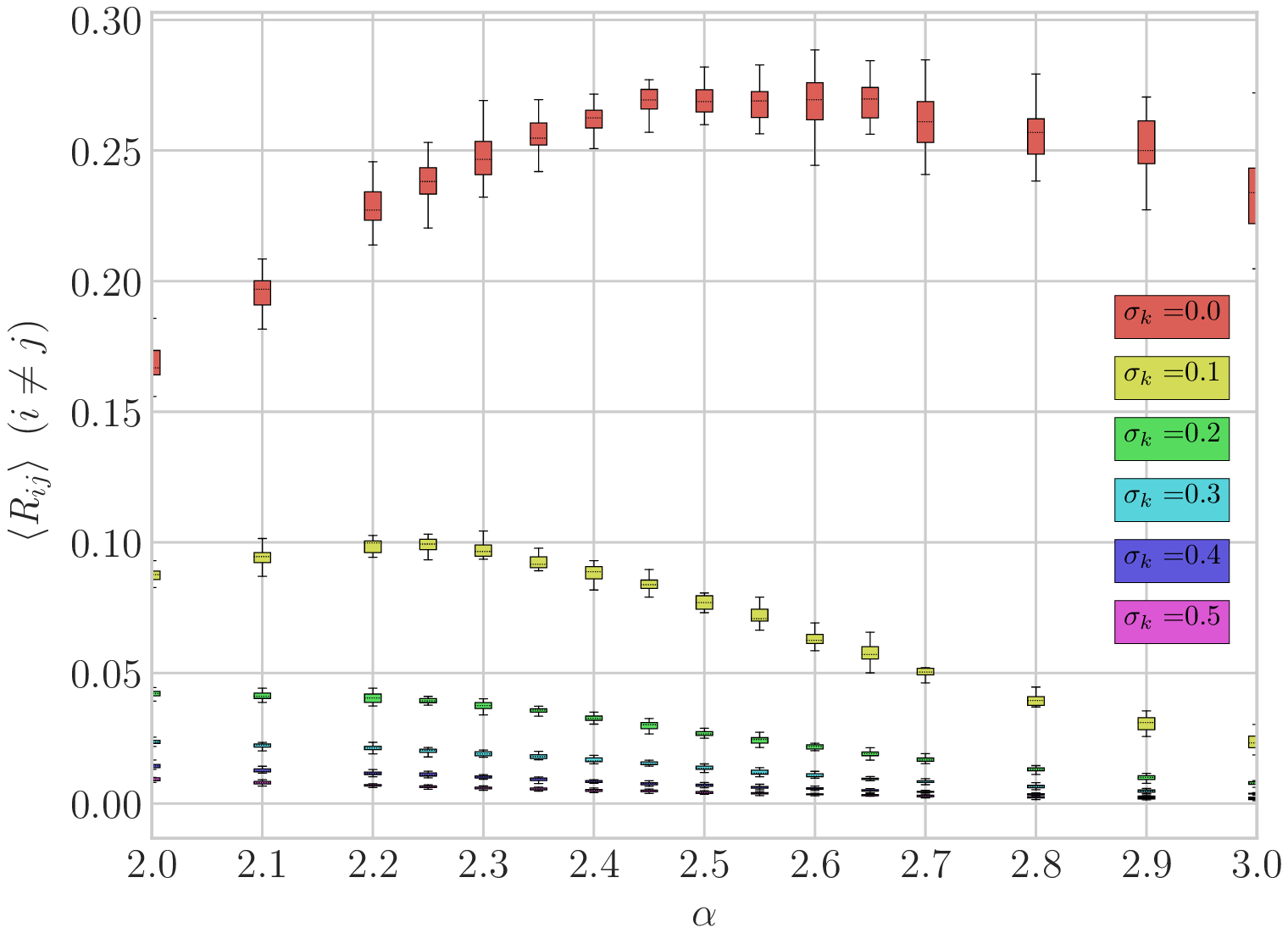}
	\label{R_heterogeneous}
	\end{subfigure}
	\hfill
	\begin{subfigure}[c]{0.8\textwidth}
	\centering
	\includegraphics[width=	0.66\linewidth]{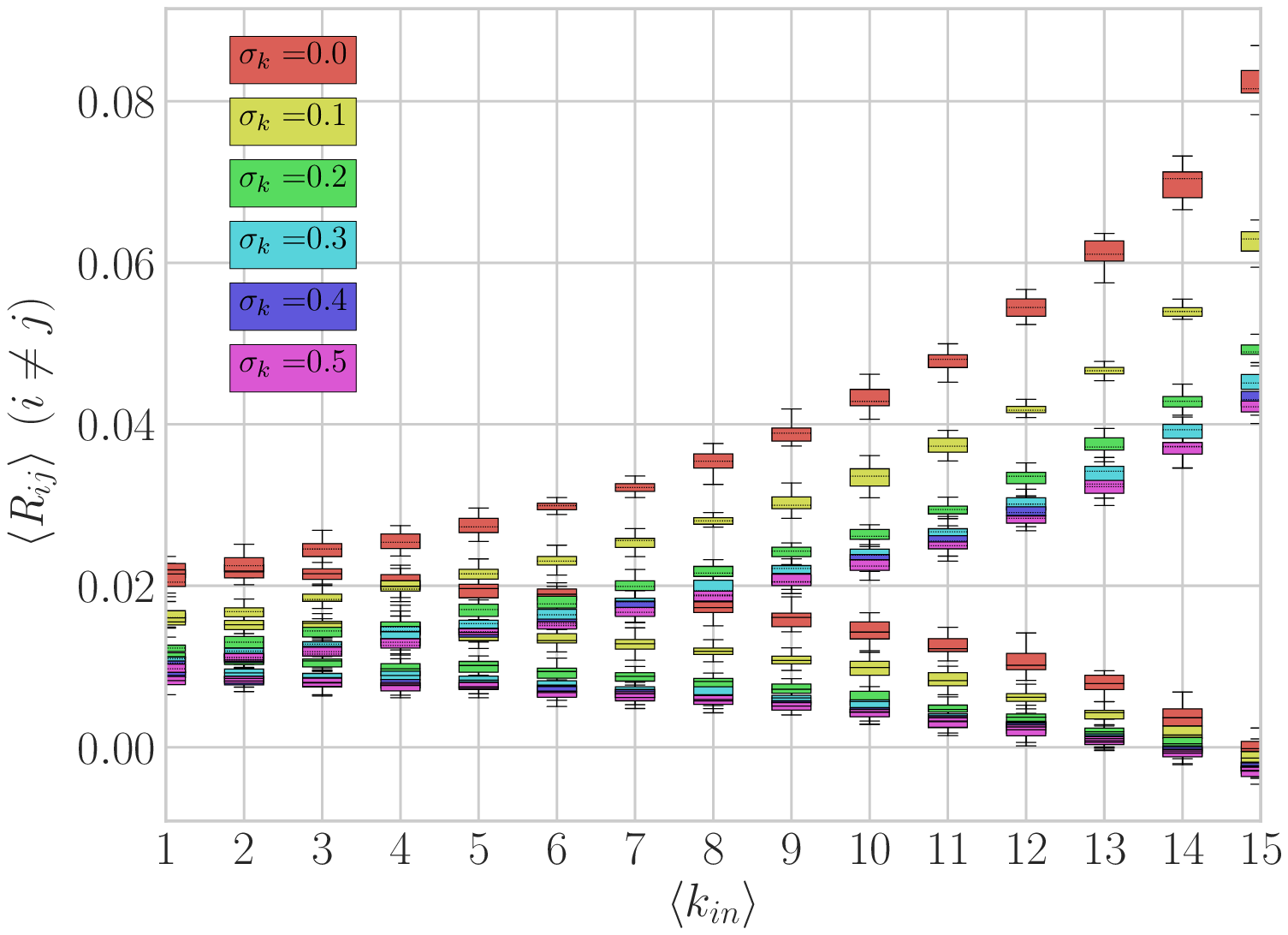}
	\label{R_newman}
	\end{subfigure}
\vspace{-0.5cm}
\caption{\small{(Color online) For model A, defined in (\ref{modelA}), the average correlation between variables, $\left\langle \mathit{R}_{ij} \right\rangle$ is plotted as a function of the characteristic network parameter, $\{\mu \}$, for 3 network topologies: Erd\"{o}s-R\'{e}nyi of size $N=50$ (upper left), heterogeneous network of size $N=200$ (upper right) and Newman modular of size $N=256$ and $\left\langle k_{total} \right\rangle=16$, where intracluster (dimmer upper) and intercluster (solid bottom) correlations are held separately (bottom). The distribution of each value is captured by a boxplot computed from 50 independent realizations. 6 different values of $\sigma_K$ account for the variability of $K$ parameter. As $\sigma_K$ increases, $\left\langle \mathit{R}_{ij} \right\rangle$ is scaled down, being the effect much larger in heterogeneous networks. In the case of ER network, $\left\langle \mathit{R}_{ij} \right\rangle$ grows with increasing $p$ until metric state eventually becomes unstable and nodes are gradually absorbed by optimal state. Larger dispersion on $K$ shifts the peak towards lower values of $p$. Conversely, for an heterogenous network, an increase on $\alpha$ exponent leads to more stability of metric state, although stable conditions are more fragile. Larger dispersion on $K$ shifts the peak towards much lower values of $\alpha$. Finally, in the case of Newman modular network, stability conditions of metric state, (\ref{stability_metric}), are always true for this parameters and hence the absence of the peak. For a better understanding of change of stability landscape, see Figure \ref{fig:metric_stability}. The parameters of the model are set as Figure \ref{correlations_histogram}.}}
\label{fig:correlations_evolution}
\end{figure}

Figure \ref{fig:correlations_evolution} not only captures the positive manifold, but also the effect of network topology on the correlations. $\left\langle \mathit{R}_{ij} \right\rangle$ is indeed modulated by the characteristic parameter of the network, $\{\mu\}$. For  Erd\"{o}s-R\'{e}nyi and Newman modular networks $\left\langle \mathit{R}_{ij} \right\rangle$ peaks around the transition between metric and optimal stable states, with a profile shaped by the network topology. Average correlation increases as connectivity does so, until nodes are recurrently being absorbed by optimal state and, consequently, $\left\langle \mathit{R}_{ij} \right\rangle$ is scaled down. Namely, Erd\"{o}s-R\'{e}nyi network peaks around $p_C$, that is when stability changes landscape. As $p$ increases more and more variables reach the optimal value and become independent from others. The impact of more variability in $K$ parameter across variables is the decrease on the correlations and the peak shift towards lower values of $p$, since stability condition (\ref{stability_metric}) is more likely to be broken. The peak generated by an heterogeneous network is more diffuse, owing to a wider distribution of the spectra. Moreover, the effect of increasing variability on $K$ has a stronger attenuation effect. On the contrary, the absence of a peak coming out from Newman modular network is explained in Section \ref{stability_conditions}. Nevertheless, a clear pattern emerges if we split intracluster from intercluster mean correlations. While intracluster correlations increase as  $\left\langle k_{in} \right\rangle$ does so, intercluster correlations decrease. However, the total mean correlation remains unchanged, as long as condition (\ref{stability_metric}) is true.

For model B, $\left\langle \mathit{R}_{ij} \right\rangle$ continuously increases as connectivity does so, until divergence conditions are met, and solutions do not longer exist. 

\subsection{From correlations matrix to statistical models}
\label{statistical_models}
When it comes to describe variability among observed, correlated variables, a bunch of statistical models comes along, aiming to approximate and understand reality. Factor analysis is a statistical method developed and widely used in psychometrics \cite{Cattell1978,McAuley1989,Barratt1965,Cohen2009}, inter alia. Observed variables are described as linear combinations of unobserved latent variables or factors, plus individual error terms, such that covariance or correlation matrix may be explained by fewer latent variables.

Historically, the most widely held theories of cognition and intelligence are built upon factor analysis raising from large batteries of conducted psychometric tests. There is still no agreement on the proper model and the underlying process which bring about the observed outcome. We list the models stood for the most outstanding theories: one factor models, multifactorial models, hierarchical models and other more complex structural models \cite{Kline1998}.

Section \ref{generating_correlations} evinces that connectivity structure between variables, mapped on a particular network topology, gives rise to different correlation matrices, even no explicit constrains on correlations between parameters are imposed. We show that certain factor models can be explained by a mutualistic dynamical model running on a particular network of variables. 

In order to avoid subjective criteria when deciding the number of factors, several methods have been developed: Horn's parallel analysis, Velicer's map test, Kaiser criteria, Cattell scree plot or variance explained criteria \cite{Cattell1966,Horn1965,Zwick1982,Costello2005}. We proceed to compute the scree plot so as to obtain the number of main principal components and hence potential latent factors. Thereupon, the statistical significance of the factor model is assessed by means of the p-value, which enables us to explore the effect of changing the parameter which best characterizes the network, $\{\mu\}$.

\begin{figure}[h!]
\centering
\includegraphics[width=	0.8\textwidth]{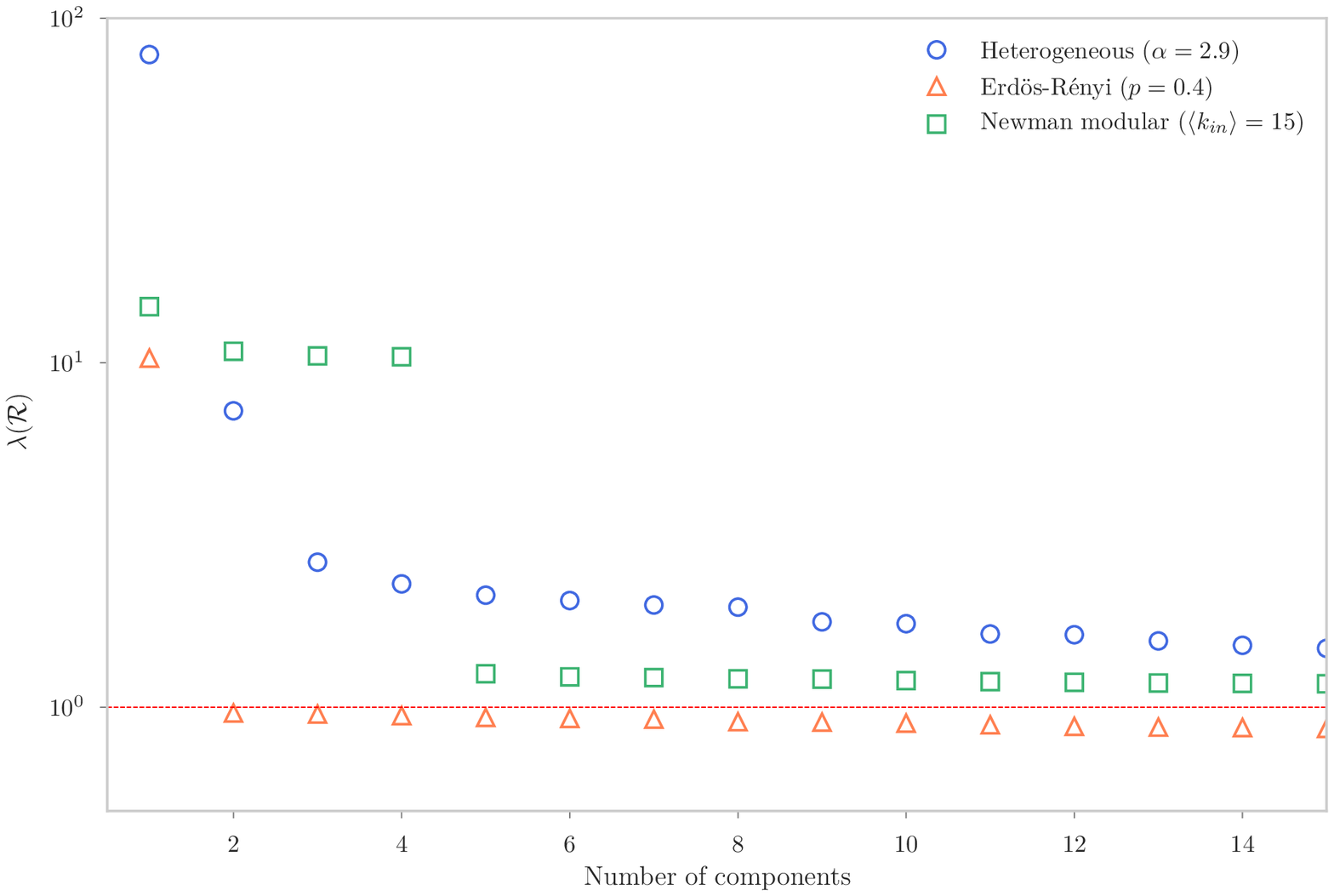}
\caption{(Color online) Scree plot, in logarithmic vertical scale, calculated from the correlation matrix, $\mathit{R}$, for 3 network topologies: Erd\"{o}s-R\'{e}nyi (triangular orange markers), heterogeneous (circular blue markers) and Newman modular (squared green markers) networks, with parameters as in Figure \ref{fig:correlations_evolution}. A scree plot shows the value of eigenvalues, $\lambda(\mathit{R})$, in descending order, as components (principal orthogonal directions) are gradually being included (up to the number of variables). $\lambda = 1$ is labelled in red, as a possible selection criterion for the number of components or factors. Other used criteria: $\lambda(\textit{R})$ much larger than the others and describing rather straight angles with successive values. 1-factor model and 4-factors model may be suitable for ER and Newman modular networks, respectively, whereas a more complex model is needed for heterogeneous network.}
\label{scree_plot}
\end{figure}

In Figure \ref{scree_plot} we have selected one specific network for each topology, given by the value of the characteristic parameter. The choice is such that solution is given by metric stable state (\ref{modelA_fixed_K}).

The number of retained components, or factors, can be obtained according to different criteria. We look at values which are much larger than 1 and describe rather straight angles with the successive values.

With these criteria and from Figure \ref{scree_plot}, we hypothesize that the correlation matrix of variables which are connected following an Erd\"{o}s-R\'{e}nyi network are well described by a 1-factor model. In the case their connectivity structure is better captured by a Newman modular network with $n$ clusters, then an $n$-factor model can not be rejected. Conversely, considering an heterogeneous network, a factorial model is no longer proposed, as eigenvalues are not clearly separated according to former criteria, but rather they follow a smooth decreasing curve. Alternatively, a statistical model which accounts for hierarchy between variables or more complex structural modelling and path analysis may be more realistic. However, this latter analysis is out of the scope of this paper.   

For the cases of Erd\"{o}s-R\'{e}nyi and Newman modular networks with 4 clusters we show that a 1-factor model and a 4-factor model, respectively, can be accurate models to obtained results. To do so, we compute the p-value of such models for different numbers of retained factors. p-value in this case is testing the hypothesis that the model fits the results perfectly and hence, we seek values $>0.05$.

In the case of an Erd\"{o}s-R\'{e}nyi network, $p_{value} \approx 1 \gg 0.05$ for $n_{factors} \geq 1$ and hence, a 1-factor model is highly likely. Similarly, in the case of a Newman modular network with $n=4$, $p_{value} > 0.05$ only when $n_{factors} \geq 4$, as suggested.

Once we have figured out the number of factors with respect to each network, we explore the effect of connectivity on the reinforcement of the statistical models, by looking at the explained variance of the most important components:
\begin{equation}
\label{prop_variance}
Var(\mathit{R})_i = \frac{\lambda_i}{\sum_k {\lambda_k}}
\end{equation}
We make use of (\ref{prop_variance}), which gives the proportion of explained variance of $\mathit{R}$ correlation matrix for each principal component of PCA. Although PCA and FCA are not equivalent statistical models \cite{Fabrigar1999,Ritter2012}, after having sustained the validity of a factor model, the former approach is acceptable in order to justify the suitability of the number of factors or components considered.
\begin{figure}[h!]
\centering
	\begin{subfigure}[t]{0.48\textwidth}
	\includegraphics[width=	1.1\linewidth]{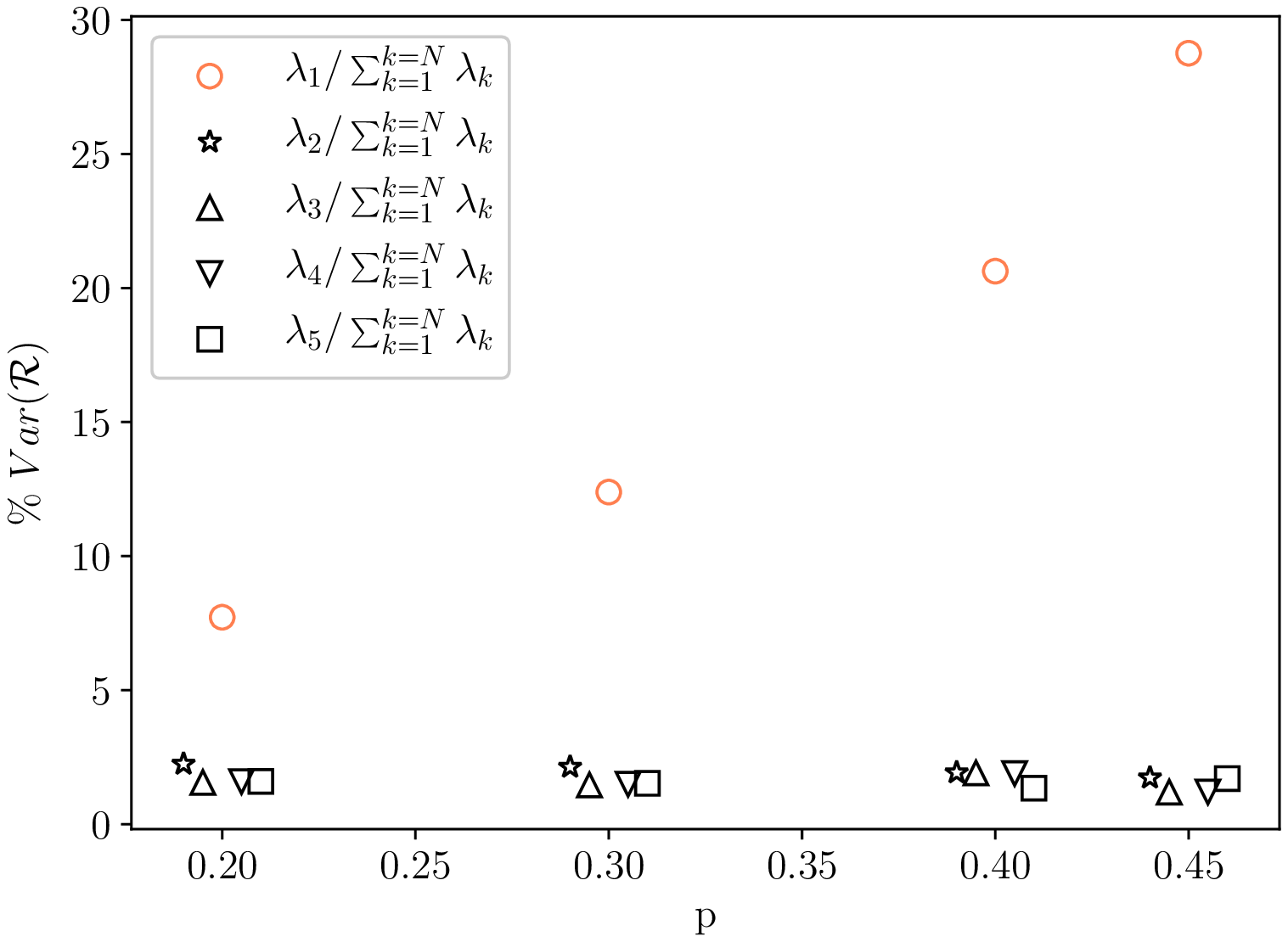}
	\label{variance_ER}
	\end{subfigure}
	\hfill
	\begin{subfigure}[t]{0.48\textwidth}
	\includegraphics[width=	1.1\linewidth]{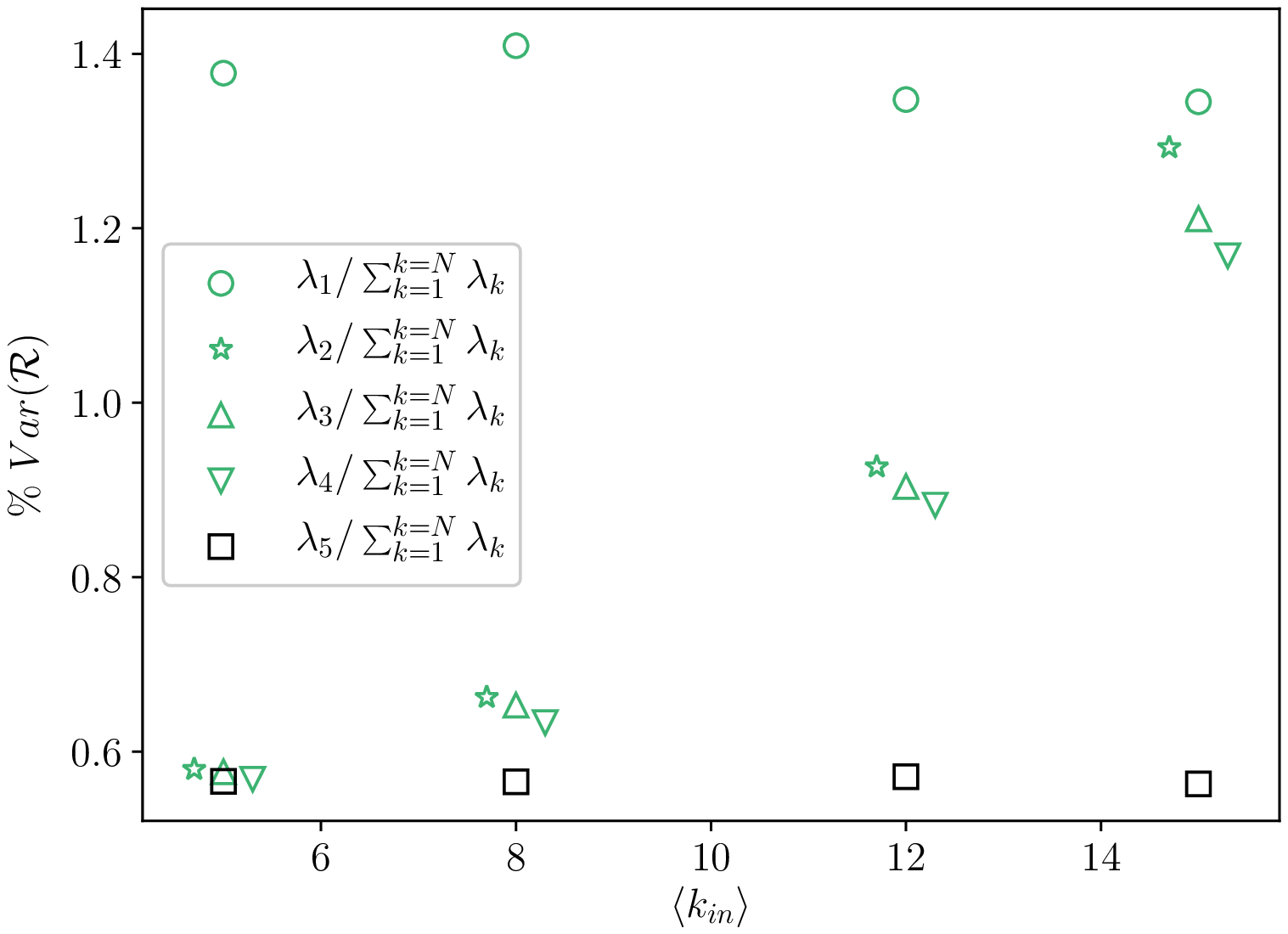}
	\label{variance_newman}
	\end{subfigure}
\vspace{-0.5cm}
\caption{\small{(Color online) Percentage of the total explained variance of correlation matrix $\% Var(\mathcal{R})$, (\ref{prop_variance}), for each of the first five components ($\lambda(\textit{R})$) as a function of the edge probability, $p$ for an Erd\"{o}s-R\'{e}nyi of size $N=50$ (left) and intracluster degree, $\left\langle k_{in} \right\rangle$ for a Newman modular network of size $N=256$ and $\left\langle k_{total} \right\rangle=16$ (right). Both networks enable a factor model as a good descriptor of the outcome. Selected significant number of components are highlighted in orange and green colors, for ER and Newman modular networks, respectively. As nodes become more connected and communities more delimited factor model moves from mirroring a noisy identity matrix to be a clear indicator of the correlation structure. First component (circular orange marker) increases as $p$ does so, strengthening the validity of a 1-factor model (left), while second to fourth components (circular, starry and triangular green markers) increase as $\left\langle k_{in} \right\rangle$ does so. Hence, an incrase on modularity reinforces the validity of a 4-factor model (right). Parameters are set as in Figure \ref{fig:metric_stability}, constraining the values to lay within the metric stable state regime.}}
\label{fig:variance}
\end{figure}

Figure \ref{fig:variance} confirms the assumptions underlying the models for both networks: in the case of an Erd\"{o}s-R\'{e}nyi network, the first eigenvalue increases as connectivity, characterized by $p$, does so and hence, a 1-factor model is being reinforced. Similarly, in the case of a Newman modular network, besides the first eigenvalue, which is always large, second to fourth eigenvalues increase as modularity, characterized by $\left\langle k_{in} \right\rangle$, does so. In contrast to eigenvalues on further positions, which remain unchanged or smaller.

Covariance matrix can be analytically computed in the case of a complete network and an heterogeneous parameter landscape, $K \sim N(\mu_K, \sigma_K)$ (Appendix \ref{appendix:covariance}).
\section{Conclusions}
\label{conclusions}
Despite mainstream approaches to cognition and intelligence research are built on static and statistics based models, we explore the emerging dynamical systems perspective putting a greater emphasis on the role of the network topology underlying the relationships between cognitive processes.

We go through a couple of models of distinct cognitive phenomena and yet find the conditions for them to be mathematically equivalent. Both models meet the requirements set out by empirical observation and established theories regarding the corresponding cognitive phenomena to which they aim to provide an explanation. Furthermore, the applied mathematical formulation may well enlight models of many real mutualistic systems, other than cognitive.

The topology of the network defined by the dynamical influence between processes indeed underlays further analysis of the results. We find the principal attractor of the system to be the exact definition of Katz-Bonacich centrality, a measure of a node importance which can also be understood as a non-conservative biased random walked along a network. We propose that heterogeneities in the dynamical parameters can be absorbed by a rescaling of the adjacency matrix weights and hence leading to the same result.

Individuals may differ in the genetic-environmental markers captured by the parameters of the model, but also in the connectivity structure between brain regions, either structural or functional. Two individuals might achieve the same performance through different neuronal routes and cognitive strategies when solving cognitive tasks. Although certain brain structures and functional pathways may be more likely to be involved in intelligence than others, there is also considerable heterogeneity which can lead to similar outcomes \cite{Deary2010}. For instance, as Newman modular network increases its modularity, the corresponding correlation matrix becomes more and more likely to be well described by a factorial model with as many factors as communities has. However, whereas the inner structure gradually changes, the average result of its stable state remains unchanged. 

The connectivity structure between cognitive processes is not known but yet it is not any. We show that network topology by its own leads to different plausible statistical models. Regardless of the network considered, it is always possible to set a parameter configuration such that the positive manifold results from the dynamical model. However, the correlations structure is determined by the network topology. Complete and Erd\"{o}s-R\'{e}nyi networks are constrained to bring about a one-factor model, more clearly defined as connectivity increases. Newman modular network enables higher order factor models, depending on the number of defined communities. Latent factors turn to be more distinguishable as modules grow to be more isolated. Conversely, heterogeneous networks lead to more complex statistical models, namely with richer and more correlated structures.

In the present article we exploit the interplay between the dynamics and the underlying network topology to model cognitive abilities and we conclude that both of the two are relevant. Although scholars are not yet sure of the relationship between cognitive processes and of the nature of intelligence we can shed a bit of light by proposing an alternative framework which captures the real meaning of `process' and `relationship': a dynamic complex network framework to model cognition.

This work is an open door to further research: we show that different network topologies lead to different correlation structures. Still, richer topologies can be considered and may bring about other interesting and eventually more realistic structures. We have restricted ourselves to static networks, though the more general definition of a network is not time constrained. What if the network which captures the connection between cognitive processes or brain modules was time dependent? What if cognitive processes could be modelled as a multilayer network, from generalists to specialists layers? Moreover, when looking only at the attractors of the system we are missing the temporal evolution of such processes and its real causality. Therefore, are these models able to explain evolving properties of the considered variables? We finally highlight that there exist several limitations in models based on ecologic systems, as exhaustively studied in population dynamics research, namely inconsistent results coming from unbounded models and discrepancy with the behaviour of some real systems\cite{Montoya2006,Goh1977,Feng2017,Pastor2015}. Hence, alternative mathematical models which overcome some of these problems shall be investigated.
\section*{Conflict of Interests}
The authors declare that there is no conflict of interests regarding the publication of this paper.
\section*{Funding Statement}
The authors acknowledge financial support
from MINECO under projects No. FIS2012-38266-C02-02
and No. FIS2015-71582-C2-2-P (MINECO/FEDER); and Generalitat
de Catalunya under the grant No. 2014SGR608. G.R.-T. acknowledges MECD under grant No. FPU15/03053.
\section*{Acknowledgements}
We wish to express our gratitude to Joan Gu\`{a}rdia for his revision and comments on the work.
\section*{Appendices}
\addcontentsline{toc}{section}{Appendices}
\renewcommand{\thesubsection}{\Alph{subsection}}

\subsection{Average and variance of stable solution for a complete and an Erd\"{o}s-R\'{e}nyi network}
\label{appendix:solution_ER}
\renewcommand{\theequation}{A.\arabic{equation}}
We define homogeneous configuration as follows:
\begin{eqnarray}
K_i \equiv K \ \forall i & r_i \equiv r \ \forall i & W_{ij} \equiv w \ \forall (i,j)
\end{eqnarray} 
In this case, metric stable solution (\ref{modelB_fixed_K}) for a complete network is straighforward:
\begin{equation}
\dot{x}_i = 0 \Rightarrow \left[ 1 - \frac{x^* - \displaystyle (N-1)w x^*}{K} \right] =0 \Rightarrow  x^* = \frac{K}{1-w(N-1)} \ \forall i
\end{equation}
In the case of a complete network, as solutions are exactly the same for all nodes, the variance of stable state is null.

For an Erd\"{o}s-R\'{e}nyi network, we compute the average of metric stable state:
\begin{equation}
\label{average}
\bar{x}^* = \frac{1}{N} \sum_i \left\langle \sum_j \left[ (\mathbb{I}-W^T)^{-1} \right]_{ij} K_j \right\rangle
\end{equation}
Using (\ref{katz_walks}) and considering $W$ is a symmetric matrix, though the expression is equivalent:
\begin{equation}
\label{expansion}
\left[ (\mathbb{I}-W^T)^{-1} \right]_{ij} = \left[ (\mathbb{I}-W)^{-1} \right]_{ij} = \delta_{ij}+W_{ij}+\sum_k W_{ik}W_{kj}+ \sum_k\sum_l W_{ik}W_{kl}W_{lj}+\cdots
\end{equation}
Taking the average on (\ref{expansion}):
\begin{equation}
\begin{split}
\frac{1}{N} \sum_i \sum_j \left[ (\mathbb{I}-W^T)^{-1} \right]_{ij} K_j = K \frac{1}{N} \sum_i \sum_j \left[ (\mathbb{I}-W^T)^{-1} \right]_{ij} = \\
K \frac{1}{N} \sum_i \sum_j \left( \delta_{ij}+W_{ij}+\sum_k W_{ik}W_{kj}+ \sum_k\sum_l W_{ik}W_{kl}W_{lk}+\cdots \right)  = 
\\
K \left[\frac{1}{N} \sum_i \sum_j \delta_{ij}+\frac{1}{N} \sum_i \sum_jW_{ij}+\frac{1}{N} \sum_i \sum_j\sum_k W_{ik}W_{kj}+\frac{1}{N} \sum_i\sum_j \sum_k\sum_l W_{ik}W_{kl}W_{lj}+\cdots \right]
\end{split}
\end{equation}
We proceed to calculate each term of the average:
\begin{equation}
\label{I}
\frac{1}{N} \sum_i\sum_j \delta_{ij} = 1
\end{equation}
\begin{equation}
\label{II}
\frac{1}{N} \sum_i\sum_jW_{ij} = w k
\end{equation}
where $ k \equiv \frac{1}{N} \sum_i k_i$ is the mean degree of a node.

Second and following terms account for the average number of $m$ next-nearest neighbours, denoted with $z_m$. The general expression for any network is given by \cite{Newman2002}:
\begin{equation}
\label{nn}
z_m = \left[\frac{z_2}{z_1} \right]^{m-1} z_1
\end{equation}
where $z_1 = k$.

In the case of an Erd\"{o}s-R\'{e}nyi network following a Poisson distribution, we get:
\begin{equation}
\label{nn_ER}
z_m = k^m \ \forall m
\end{equation}
Using (\ref{I}), (\ref{II}) and (\ref{nn_ER}):
\begin{equation}
\label{stable_ER}
\bar{x}^*_{ER} =  \frac{1}{N} \sum_i \sum_j \left[ (\mathbb{I}-W^T)^{-1} \right]_{ij} K_j = 1 + w k + w^2 k ^2 + w^3 k ^3 + \cdots =  \frac{K}{1-w k} \ \forall i
\end{equation}
We point out the fact that a Poisson distribution is not always a good model for an Erd\"{o}s-R\'{e}nyi network and hence (\ref{stable_ER}) is considered an approximation for $\bar{x}^*_{ER}$. 

The variance of $x^*_i$, $Var[x^*_i]_{ER}$, is a rather difficult computation and therefore we explore the behaviour at the second order of $w$, $\sim O(w^2)$:
\begin{equation}
\label{covariance}
Cov(x^*_i, x^*_j) \equiv Cov(x_i, x_j) =  \left\langle x_ i x_j\right\rangle - \left\langle x_ i\right\rangle \left\langle x_ i\right\rangle
\end{equation}
where $\left\langle \cdot \right\rangle$ corresponds to the average of the ensamble.

The second term, $\left\langle x_ i\right\rangle$ is already known (\ref{stable_ER}), as it is given by $\bar{x}^*$ in the thermodynamic limit. Using (\ref{modelB_fixed_K}), the first term can be expanded as:
\begin{equation}
\begin{split}
\label{cov_I}
\displaystyle \frac{x_i x_j}{K^2} =  \sum_q \delta_{jq}+w\sum_q A_{jq}+w^2\sum_q\sum_{k^{'}}A_{jk^{'}}A_{{k^{'}}q}+w\sum_q\delta_{jq}\sum_pA_{ip}+\\+w^2\sum_qA_{jq}\sum_pA_{ip}+w^2 \sum_q\delta_{jq}\sum_p\sum_kA_{ik}A_{kp}+O(w^3)
\end{split}
\end{equation}
Using (\ref{nn}) and splitting the cases $i = j$ and $i \neq j$:
\begin{equation}
\label{covariance_ER}
\begin{cases}
Cov(x^*_i, x^*_j) _{ER} \approx O(w^3) \ \forall i, j\\
Var[x^*_i]_{ER} \approx k^2w^2Var(k) + O(w^3) \ \forall i
\end{cases}
\end{equation}
Therefore, in order to obtain further information of the structure of the covariance and correlation matrices we ought to compute higher orders on $w$.
\subsection{Covariance matrix for a complete network and heterogeneous parameters configuration}
\label{appendix:covariance}
\renewcommand{\theequation}{B.\arabic{equation}}
Although the solution when connectivity structure is captured by a complete network is trivial, we can go a step further when taking into account variability in the dynamic parameters. Let us consider $K$ parameter taken from a normal uncorrelated distribution, $K_i \sim N(\mu_K, \sigma_K)$:
\begin{equation}
\left\langle K_i K_j \right\rangle =
\begin{cases}
\left\langle K_i \right\rangle \left\langle K_j \right\rangle = \mu_K ^2 \hspace{2cm} if \ i \neq j \\
\sigma_K ^2 + (\left\langle K_i \right\rangle) ^2 = \sigma _K ^2 + \mu_K ^2
\end{cases}
\end{equation}
If we separate $\left\langle x_i x_j \right\rangle$ according to the contribution regarding $K$ parameter:
\begin{equation}
\left\langle x_i x_j \right\rangle = a^2 \left\langle K_i K_j \right\rangle + ab \left\langle K_i\sum_{q\neq j} K_q \right\rangle + ab \left\langle \sum_{p\neq i} K_p K_j \right\rangle + b^2 \left\langle \sum_{q\neq j} K_q \sum_{p\neq i} K_p\right\rangle
\end{equation} 
where $a  \equiv \displaystyle\frac{1 - (N-2)w}{1-(N-2)w-(N-1)w^2}$ and $b \equiv \displaystyle \frac{w}{1-(N-2)w-(N-1)w^2}$. 

Using (\ref{covariance}), covariance matrix can finally be written as follows:

\begin{equation} 
\label{cov_complete}
Cov(X) \equiv
  \begin{pmatrix}
     Var(x) & Cov(x) & ... & Cov(x)\\
     Cov(x) & Var(x) & ... & Cov(x) \\
     ...&...&...&...\\
     Cov(x) & Cov(x) & ... & Var(x)

  \end{pmatrix}
\end{equation}
where $$Var(x) \equiv \displaystyle \frac{K^2[-w^2(N-2)^2+2w(N-2)-1]+\sigma_K^2[w^2(N-2)+(1-(N-3)w)]}{(1+w)^2(1-(N-1)w)^2}$$ and $$Cov(x) \equiv \displaystyle \frac{K^2 [-w^2(N-2)^2 + 2w(N-2)-1]+\sigma_K^2w^2(N-2)}{(1+w)^2(1-(N-1)w)^2}$$
From (\ref{cov_complete}) it turns out that a 1-factor model is indeed valid for the description of the covariance matrix. In cases where $Cov(x) \sim Var(x)$ then adequacy decreases. The particular case $Cov(x) = Var(x)$ is achieved when $\tilde{w} = \frac{1}{N-3}$. However, such condition is never reached: 
\begin{equation}
\tilde{w} > \frac{1}{N-1}
\end{equation}
As seen from (\ref{modelB_complete}), stable state diverges when $w > \displaystyle\frac{1}{N-1}$. 
\subsection{Stability conditions}
\renewcommand{\theequation}{C.\arabic{equation}}
\label{appendix:stability}
To obtain the stability conditions of optimal fixed point (\ref{modelA_fixed_C}), we expand (\ref{modelA}) around this fixed point $x_i \equiv x(C)_i + \epsilon _i$: 
\begin{equation}
\label{C_stability}
\dot{\epsilon}_i = r_i (C + \epsilon _i)\left(1 -\frac{(C + \epsilon _i)}{C}\right) \left[ 1 - \frac{(C + \epsilon _i) - \displaystyle {K_i}/{r_i}\sum_{j}W_{ji}(C + \epsilon _i)}{K_i} \right]
\end{equation}
If we linearize (\ref{C_stability}) keeping only terms $\sim O(\epsilon_i)$, we obtain:
\begin{equation}
\dot{\epsilon}_i \approx -\epsilon_i \left[ r_i \left( 1-\frac{C}{K_i} \right)+C \sum _j {W_{ji}} \right] \equiv -\beta_i \epsilon_i
\end{equation}
On the other hand, to obtain the stability conditions of metric fixed point (\ref{modelA_fixed_K}), we expand (\ref{modelA}) around this fixed point $x_i \equiv x(W_d)_i + \epsilon _i$:  
\begin{equation}
\label{metric_stability}
\dot{\epsilon}_i \approx -x(W_d)_i \left( 1- \frac{x(W_d)_i}{C}\right)\left(\epsilon_i - \frac{K_i}{r_i}\sum_j{W_{ji}\epsilon_j} \right)
\end{equation}
(\ref{metric_stability}) can be written in matrix forms as (\ref{S_matrix}).

We can derive a threshold for the stability condition using the Perron-Frobenius theorem, (\ref{perron-frobenius}), applied to (\ref{S_matrix}) which enables us to write:
\begin{equation}
\label{PF_metric}
\lambda_{\max}(\mathcal{S}) < \left[-x(W_d)_i \left( 1- \frac{x(W_d)_i}{C}\right)\left(1 - \frac{K_i}{r_i}\sum_j{W_{ji}} \right)\right]_{max}<0
\end{equation}
Looking into the extreme conditions of (\ref{PF_metric}) we conclude:
\begin{equation}
\left(1 - \frac{K_i}{r_i}\sum_j{W_{ji}} \right) > 0 \Rightarrow \frac{K_i}{r_i}\left(\sum_j{W_{ji}} \right)_{\max} \equiv \left(\sum_j [\mathbb{W}_d]_{ij}\right)_{\max} <1
\end{equation}
\bibliographystyle{unsrt}

\end{document}